\documentclass[preprint,journal]{vgtc}                % final (journal style)
\PassOptionsToPackage{bookmarks=false,draft}{hyperref}

\def\sheparding{}
\let\sheparding\undefined

\ifx\sheparding\undefined
\newcommand\changed[1]{{#1}}
\newcommand\updatenote[1]{}
\newcommand\deleted[1]{}
\usepackage{hyperref}
\else
\usepackage{soul}
\newcommand\changed[1]{{\textcolor{green}{#1}}}
\newcommand\updatenote[1]{{\textcolor{green}{#1}}}
\newcommand\deleted[1]{{\textcolor{red}{\st{#1}}}}
\usepackage[hidelinks]{hyperref}
\fi

\ifpdf%                                % if we use pdflatex
  \pdfoutput=1\relax                   % create PDFs from pdfLaTeX
  \usepackage{graphicx}                % allow us to embed graphics files
  \DeclareGraphicsExtensions{.pdf,.png,.jpg,.jpeg} % for pdflatex we expect .pdf, .png, or .jpg files
\else%                                 % else we use pure latex
  \usepackage{graphicx}                % allow us to embed graphics files
  \DeclareGraphicsExtensions{.eps}     % for pure latex we expect eps files
\fi%

\graphicspath{{figures/}{pictures/}{images/}{./}} % where to search for the images

\usepackage{microtype}                 % use micro-typography (slightly more compact, better to read)
\PassOptionsToPackage{warn}{textcomp}  % to address font issues with \textrightarrow
\usepackage{textcomp}                  % use better special symbols
\usepackage{mathptmx}                  % use matching math font
\usepackage{times}                     % we use Times as the main font
\usepackage{cite}                      % needed to automatically sort the references
\usepackage{tabu}                      % only used for the table example
\usepackage{booktabs}                  % only used for the table example
\usepackage[pdftex,dvipsnames]{xcolor}
\usepackage{amsmath}
\usepackage{mathtools}
\usepackage{euscript}
\usepackage{tikz}
\usetikzlibrary{topaths,calc}
\usepackage{subcaption}
\DeclareCaptionFont{8pt}{\fontsize{8pt}{9.52pt}\selectfont}
\usepackage{todonotes}
\usepackage{mathrsfs}
\usepackage{amsmath}
\usepackage{amssymb}
\usepackage[babel]{csquotes}

\definecolor{ColorTeaserLabelA}{RGB}{46,84,150}
\definecolor{ColorTeaserLabelB}{RGB}{197,90,17}
\definecolor{ColorTeaserLabelC}{RGB}{248,55,92}
\definecolor{ColorTeaserLabelD}{RGB}{230,180,40}
\definecolor{ColorDKModelChangeReject}{RGB}{255, 77, 79}
\definecolor{ColorDKModelChangeAccept}{RGB}{90, 214, 28}
\definecolor{ColorDKModelChangeNegative}{RGB}{141, 1, 81}
\definecolor{ColorDKModelChangeNeutral}{RGB}{150,150,150}
\definecolor{ColorDKModelChangePositive}{RGB}{39,100,25}
\definecolor{ColorDKModelSelection}{RGB}{255,0,0}

\definecolor{ColorWorkflowMLModel}{RGB}{75,115,173} % 218, 232, 252
\definecolor{ColorWorkflowVALoop}{RGB}{255,124,0} % 255,181,112
\definecolor{ColorWorkflowVisualization}{RGB}{200,162,49} % 255, 242, 204
\definecolor{ColorWorkflowFilter}{RGB}{152,65,61} % 248, 206, 204
\definecolor{ColorWorkflowProvenance}{RGB}{104,154,76} % 213,232,212
\definecolor{ColorWorkflowDomainExpert}{RGB}{70,70,70} % 242,242,242
\definecolor{ColorWorkflowRawData}{RGB}{130,130,130} %252,252,252

\definecolor{systemBlue}{RGB}{10,132,255}
\definecolor{systemGreen}{RGB}{48,209,88}
\definecolor{systemIndigo}{RGB}{94,92,230}
\definecolor{systemOrange}{RGB}{255,159,10}
\definecolor{systemPink}{RGB}{255,55,95}
\definecolor{systemPurple}{RGB}{191,90,242}
\definecolor{systemRed}{RGB}{255,69,58}
\definecolor{systemTeal}{RGB}{100,210,255}
\definecolor{systemYellow}{RGB}{255,214,10}
\tikzstyle{cblue}=[circle, draw, thin,fill=systemBlue!20, scale=0.8]
\tikzstyle{qgre}=[rectangle, draw, thin,fill=green!20, scale=0.8]
\tikzstyle{rpath}=[ultra thick, red, opacity=0.4]
\tikzstyle{legend_isps}=[rectangle, rounded corners, thin, fill=gray!20, text=blue, draw]

\DeclareRobustCommand\circledLetter[2]{
	\tikz[baseline=(char.base)]{
		\node[shape=circle,draw=none,fill=#1,text=white,inner sep=0.2pt] (char) {{#2}};
	}
}

\DeclareRobustCommand{\inlinesymbol}[1]{%
	\begingroup\normalfont
	\raisebox{-.2\height}{\includegraphics[height=1.5\fontcharht\font`\D]{#1}}
	\endgroup
}

\ieeedoi{10.1109/TVCG.2020.3030408}

\onlineid{1032}

\vgtccategory{algorithm/technique}
\vgtcpapertype{algorithm/technique}

\title{Visual Analytics for Temporal Hypergraph Model Exploration}

\author{Maximilian T.\ Fischer, Devanshu Arya, Dirk Streeb, Daniel Seebacher, \\ Daniel A.\ Keim, and Marcel Worring}
\authorfooter{
	\item
	Maximilian T.\ Fischer, Dirk Streeb, Daniel Seebacher, and Daniel A.\ Keim are with the University of Konstanz, Germany. E-mail: \{max.fischer, dirk.streeb, daniel.seebacher, keim\}@uni-konstanz.de.
	\item
	Devanshu Arya, and Marcel Worring are with the University of Amsterdam, The Netherlands. E-mail: \{d.arya, m.worring\}@uva.nl.
}

\shortauthortitle{Fischer \MakeLowercase{\textit{et al.}}: Visual Analytics for Temporal Hypergraph Model Exploration}

\abstract{
	Many processes, from gene interaction in biology to computer networks to social media, can be modeled more precisely as temporal hypergraphs than by regular graphs. This is because hypergraphs generalize graphs by extending edges to connect any number of vertices, allowing complex relationships to be described more accurately and predict their behavior over time. However, the interactive exploration and seamless refinement of such hypergraph-based prediction models still pose a major challenge.
	We contribute \textsc{Hyper-Matrix}, a novel visual analytics technique that addresses this challenge through a tight coupling between machine-learning and interactive visualizations. In particular, the technique incorporates a geometric deep learning model as a blueprint for problem-specific models while integrating visualizations for graph-based and category-based data with a novel combination of interactions for an effective user-driven exploration of hypergraph models.
	To eliminate demanding context switches and ensure scalability, our matrix-based visualization provides drill-down capabilities across multiple levels of semantic zoom, from an overview of model predictions down to the content. We facilitate a focused analysis of relevant connections and groups based on interactive user-steering for filtering and search tasks, a dynamically modifiable partition hierarchy, various matrix reordering techniques, and interactive model feedback. \deleted{Our approach allows users to detect patterns more rapidly and accurately.}
	We evaluate our technique in a case study and through \changed{formative evaluation} with law enforcement experts using real-world internet forum communication data. The results show that our approach surpasses existing solutions in terms of scalability and applicability, enables the incorporation of domain knowledge\changed{, and allows} for fast search-space traversal.
	With the proposed technique, we pave the way for the visual analytics of temporal hypergraphs in a wide variety of domains.
}


\teaser{
	\centering
	\includegraphics[width=\linewidth]{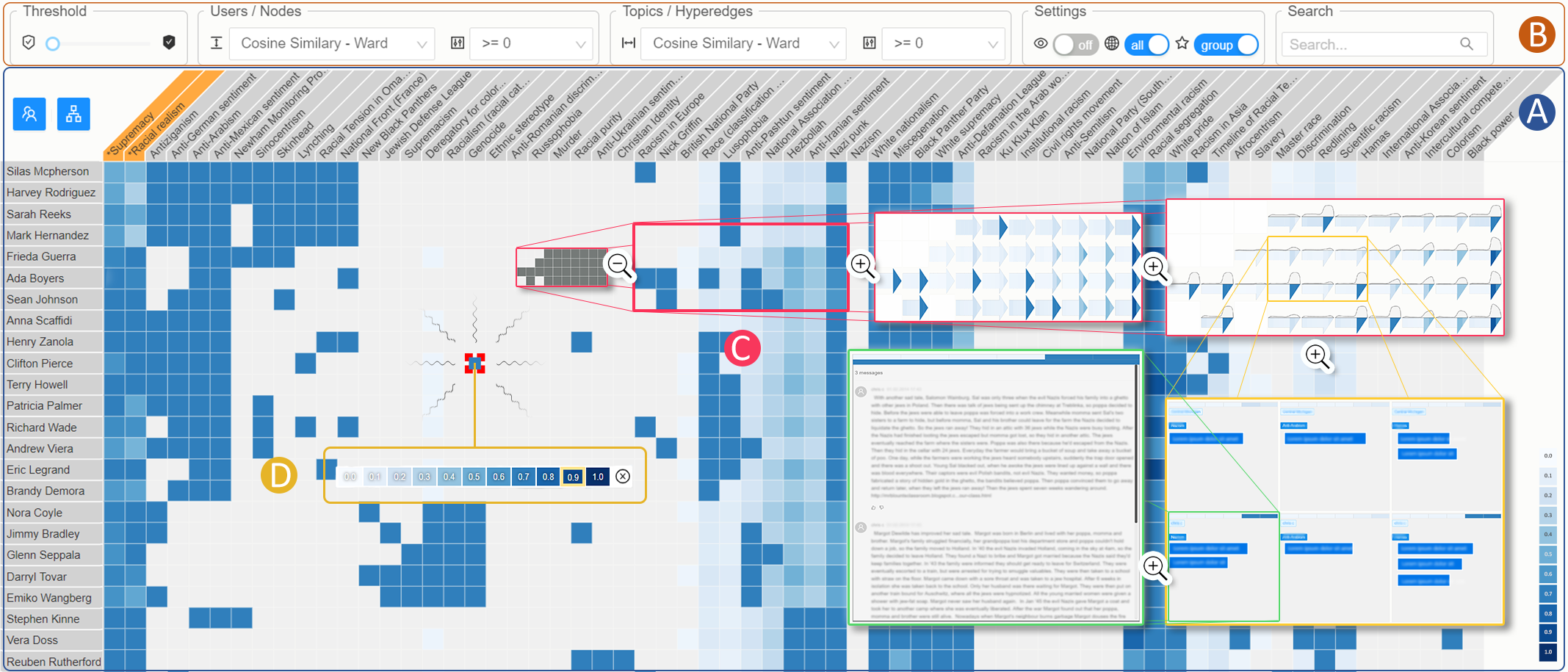}
	\caption{\updatenote{[Updated]} \textsc{Hyper-Matrix}, a novel approach to explore and refine temporal hypergraph models using visual analytics. The interactive multi-level matrix-based visualization\circledLetter{ColorTeaserLabelA}{A}enables the inspection of the model, together with the upper interface\circledLetter{ColorTeaserLabelB}{B}. The main area shows the second semantic zoom level applied to an obfuscated real-world dataset in criminal investigations, while the five insets\circledLetter{ColorTeaserLabelC}{C}show the other drill-down levels for exploration. The technique allows to interactively\circledLetter{ColorTeaserLabelD}{D}contribute domain knowledge, the resulting implications have ripple effects on the whole machine learning model, thereby refining it.}
	\label{fig:teaser}
}

\begin{document}
\firstsection{Introduction}
\label{ch:introduction}
\maketitle % must be after first section
A significant volume of real-world data consists of entities and their relationships and can accordingly be modeled mathematically using graph-based approaches.
Such approaches are widely applied in many domains, ranging from natural and social sciences to engineering and business.
Examples include modeling biological and chemical processes like protein-protein interactions~\cite{Przulj.PPIGraphModeling.2011}, relationships in computer~\cite{Wang.NetworkSecurityGraphModel.2007} as well as human communication networks~\cite{Onnela.HumCommGraph.2007}, or knowledge network exploration in business processes~\cite{Helms.KnowledgeNetworkAnalysis.2005}.
Whereas static graphs can represent the fixed relationships between entities, using an undirected or directed graph as a model, many of the examples presented above are more accurately described as processes with complex interrelations that may change or evolve.
\changed{Here, geometric deep learning methods together with interactive visualization can help to more accurately model, predict, and explore the model evolution.}
Considering, for example, conversations, a topic is a time-dependent grouping encompassing users, which cannot be described using a static graph.
This evolution of relations should be modeled by dynamic networks.
Compared to regular graphs, using edges or separate node types, such modeling often reflects the actual process more accurately.
Dynamic networks are, however, more challenging to model and have traditionally been modeled as regular, undirected graphs, mainly due to computational and visualization limitations.
In recent years, modeling has extended to dynamic networks~\cite{Kuhn.DynamicNetworks.2011}, but some limitations remain.

% using temporal hypergraphs instead
Consequently, one can take a step further and use temporal hypergraphs. 
Hypergraphs generalize graphs by extending edges to connect any number of vertices, allowing complex relationships to be described more accurately~\cite{Shi.DynamicNetworkModeling.2015} while reducing ambiguity and network inflation.
% problems with temporal hypergraphs
Utilizing temporal hypergraph prediction models, however, introduces its own set of challenges.

% 1st
First, as the model structure is more complex, it is relevant how the information is communicated to the analyst through visualization (cf.~\cite{Heintz.BeyondGraphs.2014}) \changed{and how domain knowledge feedback is incorporated}.
Static hypergraphs can be considered as standard sets, with different visualizations available~\cite{Alsallakh.SetVis.2014}.
Temporal hypergraphs, meanwhile, add a time-dependent evolution, making it harder to convey the relevant information meaningfully. 

% 2nd
Secondly, many traditional graph-based concepts cannot directly be applied to hypergraphs. 
Hyperedges, as arbitrary sized sets of connected nodes, add another order of complexity. 
In previous works~\cite{Arya.GeometricDeepLearningHypergraphs.2018, arya2019predicting}, we presented how geometric deep learning can be applied to hypergraphs and showed how this method could be leveraged to predict behavioral patterns in social media hypergraph models.

% Consequence
Consequently, the incorporation of machine learning techniques into an interactive model \changed{to more accurately predict changes in the hypergraph due to changes in the data} introduces new problems. 
While deep learning avoids assiduous manual feature engineering and algorithm design, it reduces explainability and accountability of the results. 
Domain experts usually have some domain-specific intuition---a mental model and structure---about inherent and implicit relations and groupings not available in the data, enabling them to judge the plausibility of hypotheses and to steer the exploration. 
Yet, they face difficulties articulating \changed{their domain knowledge} through machine learning into the predictions and tracing its influence. 
This holds especially for very complex models, like temporal hypergraphs. 
The knowledge formalization requires a very detailed a priori understanding of the problem by domain experts, which is not always available. 
For the same reason, it is challenging to capture the knowledge independently of the model without rapid, iterative feedback. 
Hence, the machine learning outcome often correlates strongly with the adequacy of the initial problem modeling and the quality of the training data, while domain expertise \changed{and domain knowledge are} frequently not leveraged to \changed{their} full potential.

% gap and contribution
To address these issues, we present \textsc{Hyper-Matrix}, making the following contributions:
\begin{itemize}
	\setlength\itemsep{0.0em}
	\item A novel, interactive framework for temporal hypergraph exploration through the use of semantic zooming relying on a multi-level matrix-based approach and various exploration concepts.
	\item The extension of a geometric machine learning architecture~\cite{Arya.GeometricDeepLearningHypergraphs.2018, arya2019predicting} with a relevance feedback model.
	\item A tight coupling between the visualization and the machine learning relevance feedback model for evaluation and seamless refinement, offering the integration of domain knowledge and making the corresponding model changes visually transparent.
	\item \changed{One case study} describing an application of the technique to the law enforcement field.
	\item A \changed{formative evaluation} with law enforcement experts using real-world communication data, demonstrating that our technique surpasses existing solutions, enabling the effective analysis of large amounts of information in a targeted way.
\end{itemize}
% end with a message to be remembered
Our approach bridges the gap between visual exploration and separate model training, allowing domain experts to enhance the machine learning predictions with implicit domain knowledge in the same step as evaluating and exploring the temporal hypergraph model predictions.

%
% Related Work
%
\section{Related Work}
\label{ch:related_work}
This research is an entry into the interactive temporal hypergraph model exploration in the context of explainable support by machine learning. 
In the literature hypergraphs are studied from both a visualization as well as a machine learning perspective. 
In the following discussion, we adhere to the same distinction and relate our work to the visualization of temporal hypergraphs as well as their application in machine learning.

\subsection{Visualization of Hypergraphs}
We first shortly discuss the situation for (static) hypergraphs as well as dynamic graphs, before looking at temporal hypergraphs.
Hypergraphs can be considered as a set of sets. 
The survey on set visualizations by Alsallakh et al.~\cite{Alsallakh.SetVis.2014} shows that several visualizations are applicable to hypergraphs. 
Hypergraphs are often drawn as regular graph networks or bipartite networks. 
When making their dimensionality explicit, they can be drawn as subsets---like Venn diagrams or radial sets---or in node-link form\cite{Vehlow.Survey.VisGroupGraphs.2017}, using colored hulls or other, specifically adapted approaches~\cite{Meulemans.KelpFusion.2013}. 
A third possibility is to use a matrix-based approach, which improves scalability~\cite{Kim.SetConcordance.2007}. 
Subsets and node-link diagrams suffer from limited scalability, quickly leading to occlusion and clutter.
Bound in the number of visual attributes they can employ, these techniques typically reach their constraints in the order of one or two dozens of hyperedges\cite{Valdivia.ParallelAggOrdHypergraphVis.2019}.
Further, they are difficult to extend with a temporal component, having already used up most visual attributes.

In comparison to set based approaches, dynamic graphs change over time, leaving the choice~\cite{Beck.Survey.DynamicGraphsVis.2017} between employing animation or an additional timeline component.
The former puts significant strain on the mental map when many connections change, while the latter is limited by the available screen space in the number of discrete timesteps it can show.
The survey~\cite{Beck.Survey.DynamicGraphsVis.2017} also points out that node-link diagrams remain the most commonly used type of visualization.
However, these approaches mostly lack the extendability to hypergraphs.

When studying temporal hypergraphs, the issues arising from the dimensionality and the temporal nature all build up.
Indeed, there is almost no prior work on the visualization of temporal hypergraphs specifically.
Two notable exceptions exist, which allow visualizing---but not modifying or refining---temporal hypergraphs: First, the recent works by Valdivia et al.~\cite{Valdivia.Hypenet.2017, Valdivia.DynamicHypergraphs.2018, Valdivia.ParallelAggOrdHypergraphVis.2019}. 
Their visualization approach is also shown later in \autoref{fig:comparison_paohvis} as part of the case study. 
Second, the previous work by Streeb et al.~\cite{Streeb.VAHypergraphPrediciton.2019} introduces an in-line visualization of the temporal evolution.
Valdivia et al.\ begin to tackle the research gap by proposing PAOHvis, thereby claiming to provide the \enquote{first [\ldots] highly readable representation of dynamic hypergraphs}.
While this is a strong claim to make, the literature review showed a broad diversity between the approaches, but none---except the two mentioned above---is directly suitable for temporal hypergraph visualization, supporting this conclusion.
Utilizing the previously discussed approaches as substitutes for a tailored visualization often does not adequately leverage the additional information available with temporal hypergraphs\deleted{introducing additional complexity,} and \changed{does not address} the tasks that come with hypergraph topology and evolution. \changed{For those, we refer to Section}~\ref{sec:tasks}.
\changed{Shortcomings in existing approaches include, for example, Streeb et al.\ providing only the prediction abstraction level in their visual interface (cf.~Level~3 in Section~\ref{sec:model_visualization}).  Similarly, this is true for Valdivia, although they support coloring by a group. This can lead to information overload, as filtering using thresholds is the only way to reduce the information. In contrast, usage of semantic zoom enables an exploration of the complete hypergraph (cf.~Section~\ref{sec:model_visualization}) without the need to preliminary apply filters while enabling tailored visualizations showing detailed information when focusing on different abstraction levels. Prominent examples of matrix-based visualizations are the Zoomable Adjacency Matrix Explorer~\cite{Elmqvist.ZAME.2008} that enables users to zoom and pan with interactive performance from an overview to the most detailed views and the visual analysis system of Behrisch et al.~\cite{Behrisch.VAMatrices.2014}. It features a flexible semantic zoom to navigate through sets of matrices at different levels of detail. Further, both Streeb and Valdivia, only support sorting by weights and average (cf.~size ordering in Section~\ref{sec:matrix_reordering}), compared to our default matrix-based sorting, improving cluster identification. Significantly, all existing approaches aim at analyzing a fixed hypergraph model. None focus on interactively working \emph{with} the model and iteratively improving it (cf.~Sections~\ref{sec:model_relevance_feedback} and~\ref{sec:model_updates}).}

At last, while not strictly related to the research on temporal hypergraphs \textit{per se}, we want to mention approaches that are, at least partly, similar to ours, and also conventional tools so far applied in practice.
Here we concentrate on how hypergraph-like data is handled in the law enforcement field, relevant for the case study and the evaluation through domain experts (see also Sections~\ref{ch:case_study} and~\ref{ch:domain_expert_asessment}).
The visual analysis of communication data---but without any hypergraph visualization or a tunable model---is not novel and has been researched both from the analytical side~\cite{Luo.SNACommChar.2015} as well as the visualization side~\cite{Wu.OpinionFlow.2014}. 
Also, the idea of semantic zooming for matrix-like visualizations has been described previously~\cite{vanHam.MultiLevelMatrices.2003}, however, in a different way and in the area of software management.
Further, it was also described how an overlay magic lens~\cite{Ghoniem.VAFLESemanticLense.2014} can be used instead of zooming, to keep the context and allow for faster search space traversal from locations far apart, which we partly employ for the partition hierarchy (Section~\ref{sec:partition_hierarchy}).
In practice, for the law enforcement field, we found that data which benefits from a hypergraph modeling, like communication patterns or process analysis, is prevalent, but not supported by any system. 
Gephi~\cite{Bastian.Gephi.2009} is sometimes used, but analysts often prefer Pajek~\cite{Batagelj.Pajek.1998, Batagelj.Pajek.2002}, as it supports larger networks. 
The most popular tool is IBM i2 Analyst's Notebook's~\cite{IBM.AnalystsNotebook} graph component due to the prevalence and familiarity in this domain.

% ML hypergraphs
\subsection{Machine Learning for Hypergraph Models}
Learning with hypergraphs was introduced by Zhou et al.~\cite{zhou2007learning} to model high-order correlations for semi-supervised classification and clustering. 
It generalizes the efficient methodology of spectral clustering to hypergraphs by proposing a label propagation method to minimize the differences in labels of vertices sharing the same hyperedge. 
The correlation among hyperedges was further explored by Hwang et al.~\cite{hwang2008learning}, assuming that highly correlated hyperedges have similar weights.
More recent works~\cite{bronstein2017geometric} concentrate on parametric learning of weights using propagation of node features across hyperedges~\cite{feng2019hypergraph, yadati2019hypergcn}. % bai2019hypergraph is only a prepreint so far

\par \changed{Understanding communication patterns of users on social networking sites has created opportunities for richer studies of social interactions and better prediction of behavioral patterns.} In multimedia, link prediction on hypergraphs has been a popular topic of research \changed{in social network analysis}. 
This includes predicting metadata information such as tags and groups for entities in social networks, e.g., images from Flickr~\cite{Arya.GeometricDeepLearningHypergraphs.2018}, music recommendation by exploiting network proximity information of users in Last.fm~\cite{bu2010music} and predicting higher-order links (such as tweets with a specific hashtag) in Twitter~\cite{li2013link}. Besides, hypergraph learning models are being used in multimodal data analysis to integrate complementary information from multiple modalities effectively. Liu et al.~\cite{liu2017multi} proposed a multi-hypergraph learning method to handle incomplete multimodal data for disease diagnosis in neuroimaging and Arya et al.~\cite{arya2019hyperlearn} proposed a framework to learn a compact representation for each modality in a multimodal hypergraph using a tensor-based representation. \changed{These works have shown the importance of hypergraph based learning for predicting implicit links within a network.} However, none of these approaches pose an interactive learning formulation that can assimilate user feedback as an external source of information to either improve the predictive capability of a model or to even change the intrinsic properties such as learnable parameters of a model. \changed{In this work, we extend our previous work \cite{arya2019predicting} on link prediction in communication networks capable of fine-tuning the trained model by incorporating external relevance feedbacks.}

\subsection{Tasks for Evaluation of Temporal Hypergraph Models}
\label{sec:tasks}
Tasks in temporal hypergraph analysis relate to dynamic networks and set comparisons. 
A task taxonomy of the former is provided in the survey by Beck et al.~\cite{Beck.Survey.DynamicGraphsVis.2017}, and for the latter in the survey by Alsallakh et al.~\cite{Alsallakh.Survey.SetVisualization.2016}.
For temporal hypergraphs, in particular, the tasks sometimes substantially differ; for example, one being the analysis of changes of both connections and attributes over time. 
The proposed technique does not directly fit with any existing task taxonomy, positioning itself between disciplines~\cite{Andrienko.ExploratoryAnalasis.2006}.
For a discussion on existing taxonomies and their applicability to temporal hypergraphs, we refer to the existing work by Valdivia et al.~\cite{Valdivia.ParallelAggOrdHypergraphVis.2019} and summarize only the main aspects here. 
Our technique supports not all traditional tasks in set analysis~\cite{Alsallakh.Survey.SetVisualization.2016}, and in dynamic network analysis~\cite{Lee.TaskTaxonomyGraphVis.2006, Ahn.TaskTaxNetwEvol.2014, Bach.GraphDiaries.2014, Kerracher.TaxonomyTempGraphVis.2015}, summarized in~\cite{Beck.Survey.DynamicGraphsVis.2017}. 
However, it provides support for several additional tasks relevant to our driving application. 
These include the clustering of related groups independently of their temporal connection, the inspection of shared attributes of connections, the following of temporal evolutions, while both retaining an overview and simultaneously being able to explore details. 
In short, the experts are interested in connectivity information involving both graph topology as well as attribute values, which can be separated between time ranges.
One main requirement is the need to include external (domain) knowledge that is not directly available as raw data and includes conceptualized topics in line with their mental categorization.
These tasks are not sufficiently described or supported by existing taxonomies, as they neglect the additional complexity incorporated by hypergraphs and the domain knowledge integration.

\paragraph{}
Given the sparse research in hypergraph visualization, it is unsurprising that there is no prior work on bridging both fields; this is the gap we aim to fill: offering a technique that addresses the shortcomings discussed above, enabling the exploration and refinement of hypergraph models using interactive visualization, closing the visual analytics loop.

%
% Methodology
%
\section{Extension of Machine Learning to Hypergraphs}
\label{ch:methodology}
\begin{figure*}[ht!]
	\centering
	\includegraphics[width=.94\linewidth]{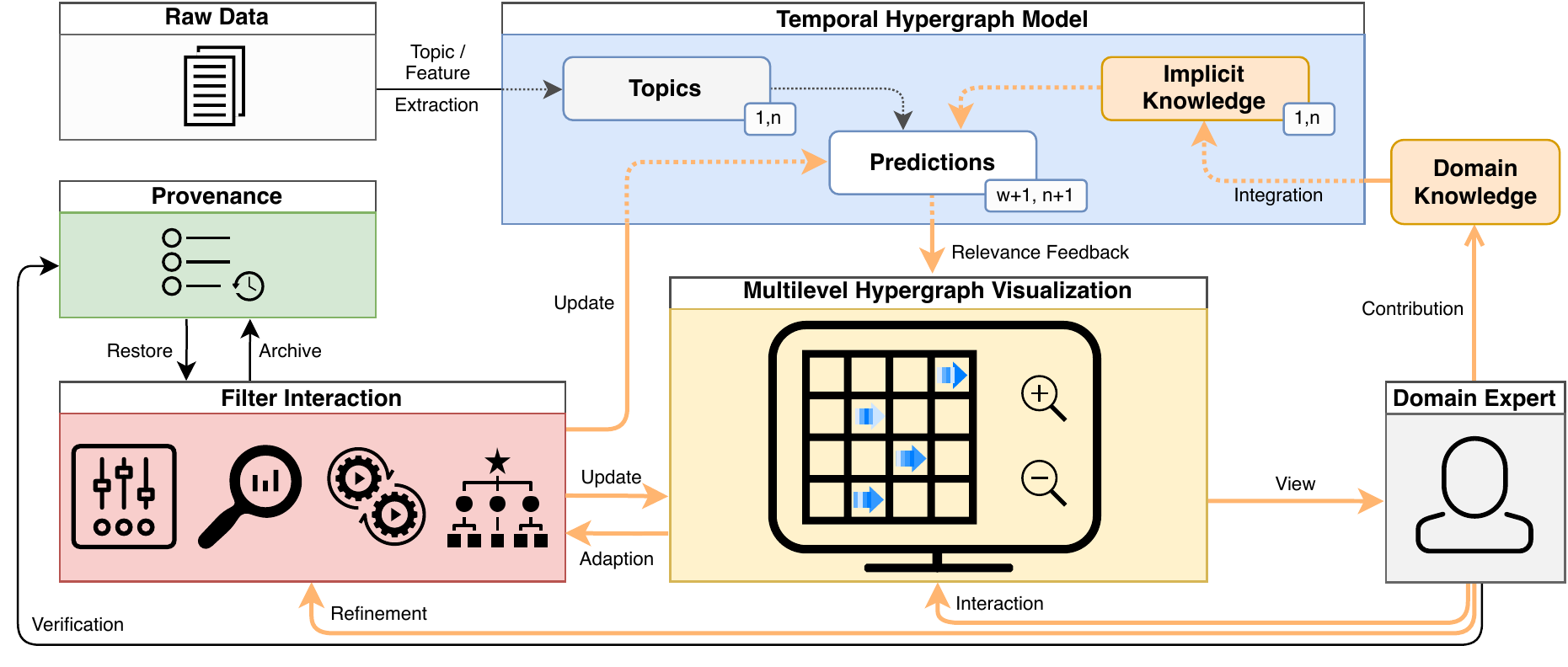}
	\caption{High-level workflow of our technique, showcasing the main components and the interaction flow for the exploration and refinement of temporal hypergraph models, adapted to use case A in Section~\ref{ch:case_study}. The workflow begins with \textcolor{ColorWorkflowRawData}{\textbf{raw data}}~\inlinesymbol{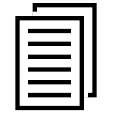}extraction and the generation of a \textcolor{ColorWorkflowMLModel}{\textbf{temporal hypergraph model}}. The model state is visualized using a matrix-based \textcolor{ColorWorkflowVisualization}{\textbf{multilevel hypergraph visualization}}~\inlinesymbol{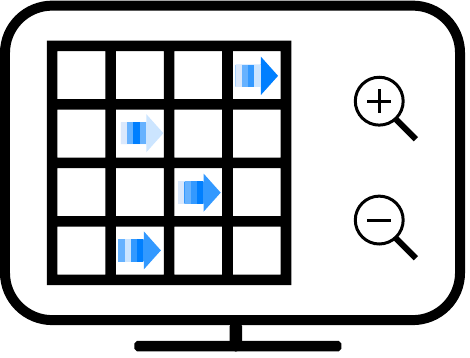}, allowing for various exploration and \textcolor{ColorWorkflowFilter}{\textbf{filter schemes}}, including search~\inlinesymbol{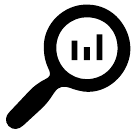} and filters~\inlinesymbol{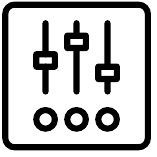}, a dynamically modifiable partition hierarchy~\inlinesymbol{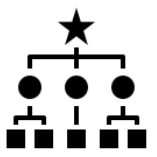}, and matrix-reordering techniques~\inlinesymbol{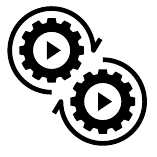}. The \textcolor{ColorWorkflowDomainExpert}{\textbf{domain expert}}~\inlinesymbol{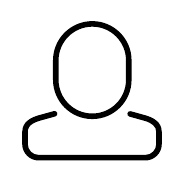}can interact with the model by either refining the filter schemes or by contributing domain knowledge, which both update the model. The model feedback can then be explored and accepted, closing the \textcolor{ColorWorkflowVALoop}{\textbf{visual analytics loop →}}. The chronology of interactions and contributions are available for recovery or verification as a \textcolor{ColorWorkflowProvenance}{\textbf{provenance history}}~\inlinesymbol{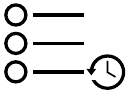}, facilitating accountability.}
	\label{fig:workflow}
\end{figure*}

In the following two sections, we describe the overall workflow of our approach, shown in \autoref{fig:workflow}.
We begin with an exemplary description of one geometric deep learning model, adapted to a task relevant for our law enforcement domain experts: the temporal prediction and analysis of patterns in communication data. 
It acts as a blueprint for problem-specific temporal hypergraph models.
In Section~\ref{ch:hypergraph_va}, we then discuss the interactive exploration  using visual analytic principles.

\subsection{Notation and Formulation of a Temporal Hypergraph}
In set theory, an undirected hypergraph $H = (V, E)$ is defined as an ordered pair, where $V = \{v_1,.., v_n\}$ represents the $n$ vertices (hypernodes) and subsets of these vertices $E = \{e_1,.., e_m\}$ constitute the $m$ distinct hyperedges. $H$ is represented by the incidence matrix $\scriptstyle \mathbb{I} = |V|\times|E|$, with entries $\scriptstyle \mathit{i}(v_j,e_k)$ $1$ if $\scriptstyle v_j \in e_k$ and 0 otherwise. \changed{We define the neighborhood of ${v_j}$ as the set $\scriptstyle \EuScript{N}(v_j)$ of nodes within the same hyperedges as ${v_j}$.}

In adapting a generic temporal hypergraph model to our use case, we follow our previous work~\cite{arya2019predicting}, representing the relationship between internet forum users and their behavioral characteristics (both ``explicit''  and ``implicit"). 
The available metadata (in particular forum category) forms the explicit characteristic of a user, while their topics of discussion outline the implicit communication characteristic. 
Thereby, we construct two separate hypergraphs depicting the connection of users with these explicit and implicit behavioral characteristics. 
To model the temporal component, let us define a temporal hypergraph by $\scriptstyle H_{[t]}$, at a given time $t$, where each user is represented as a node, and each type of explicit/implicit characteristic is represented as a separate hyperedge. We denote the explicit and implicit hypergraphs, at any given time $t$, by $\scriptstyle H^0_{[t]}$ and $\scriptstyle H^{\prime}_{[t]}$, respectively. 
Consequently, in $\scriptstyle H^{\prime}_{[t]}$, each topic is depicted as a separate hyperedge and users (nodes) who adhere to a common topic of interest are connected by it. Thus, forecasting the evolution of users' topics of interest for time $t\!+\!1$ becomes equivalent to the task of finding new relations over the existing relations in hypergraph $\scriptstyle H^{\prime}_{[t]}$.

\subsection{Relevance Feedback to Deep Learning Model}
\label{sec:model_relevance_feedback}
As indicated, the underlying model for forecasting future interests of internet forum users is based on predicting links in temporal hypergraphs. 
The task of link prediction on a hypergraph $\scriptstyle H_{[t]}$ with a fixed set of edges $E$ aims at updating the set $e_k$. 
This link prediction can be formulated as missing value imputation or a matrix completion task on~$\mathbb{I}$. 
In the following, we extend our previous work~\cite{Arya.GeometricDeepLearningHypergraphs.2018,arya2019predicting}, to allow for the incorporation of feedback in \textsc{Hyper-Matrix}. 
Therefore, we first reconsider the module for training a geometric deep-learning model.
Then, we formulate how feedback from the user can be employed to update the model.

\paragraph{Training Module} Let $\scriptstyle \mathbb{I}_{[t]}$ denote the incidence matrix of $\scriptstyle H^{\prime}_{[t]}$ at time $t$ which we can factorize as $\scriptstyle \mathbb{I}_{[t]}=X_tY_t^T$ with $\scriptstyle X_t$ and $\scriptstyle Y_t$ the row and column matrices, respectively. 
Hypergraph $\scriptstyle H^0_{[t]}$ will be utilized as an auxiliary set of explicit information between users for predicting links in the implicit hypergraph $\scriptstyle H^{\prime}_{[t]}$. 
The information in $\scriptstyle H^0_{[t]}$ is encoded by extracting its Laplacian denoted by $\scriptstyle \Delta_0$.
The Laplacian $\scriptstyle \Delta_0$ gives a measure for the relatedness between any pair of users~\cite{bu2010music}. 
Using such a similarity measure can significantly enhance the user-topic link prediction outcomes by reducing extraneous noise and thus smoothing the model output.

To train the model, \changed{we employ a semi-supervised learning setup, hence the predictive loss is backpropagated by using a small set (around 5--8\,\%) of known links in $\scriptstyle H^{\prime}_{[t+1]}$.
These known links create an upper bound for the number of timesteps the model can predict in $\scriptstyle \mathsf{\hat{\mathbb{I}}_{[t\!+\!1]}}$.
Details can be found in~\cite{Arya.GeometricDeepLearningHypergraphs.2018, arya2019predicting}.
For training,} we take the incidence matrix $\scriptstyle \mathbb{I}_{[t]}$ at time~$t$ and use the hypergraph link prediction model $\scriptstyle \mathbb{H}_{GDL}$ to learn the best parameter set $\scriptstyle \Phi[t]$ for predicting the incidence matrix $\scriptstyle \mathbb{I}_{[t+1]}$ at time $t\!+\!1$:
\begin{equation}
\mathsf{\hat{\mathbb{I}}_{[t\!+\!1]},\Phi_{[t]} = \mathbb{H}_{GDL}(\mathbb{I}_{[t]},\Delta_0)}
\end{equation}

\paragraph{Feedback Module} In order to integrate domain knowledge into the underlying model, we propose a novel interactive learning formulation to incorporate feedback from the domain expert. 
These feedbacks are assumed to contain definitive implicit information about the topic of interest for certain users in the dataset.
Instead of just updating the information by directly changing the topic (hyperedge) of the respective users (nodes), these feedbacks should also create a \changed{\enquote{ripple effect}} on the overall connections in the hypergraph $\scriptstyle H^{\prime}_{[t]}$. 
That is, if the feedback $\scriptstyle f_{[t]}$ at time $t$ involving the single user ($u_j$) denoted by node $v_j$ in the hypergraph $\scriptstyle H^{\prime}_{[t]}$, then incorporating $\scriptstyle f[t]$ will entail a twofold operation: \emph{1. Update}: Topics for user $u_j$ are updated, i.e., add/remove $v_j$ to/from the respective hyperedges $E = \{e_1,.., e_m\}$ corresponding to $\scriptstyle f_{[t]}$. \emph{2. Predict}: Change topics for users in close communication with $u_j$ based on their relatedness to $u_j$, i.e., re-calculate the connection strength for vertices in $\scriptstyle \EuScript{N}(v_j)$ with the hyperedges $E = \{e_1,.., e_m\}$.
The first operation is a straightforward updating of the matrix $\scriptstyle \mathbb{I}_{[t+1]}$ by updating new values corresponding to nodes and edges suggested in the feedbacks $\scriptstyle f_{[t]}$. 
The change in the neighborhood connections are calculated by using the updated matrix $\scriptstyle \mathbb{I}_{[t+1]}+f_{[t]}$ as input to our link prediction model $\scriptstyle \mathsf{\mathbb{H}_{GDL}}$. 
However, in the feedback module, instead of learning parameters through an iterative process, the learned parameters $\scriptstyle \Phi_{[t]}$ are used as initialization of the already trained model $\scriptstyle \mathsf{\mathbb{H}_{GDL}}$. 
This ensures the model converges in far less time after incorporating the feedbacks $\scriptstyle f_{[t]}$ than when learned from scratch. 
The following equation shows the representation of the feedback module in symbolic form:
\begin{equation}\label{equ:feedback}
\mathsf{\hat{\mathbb{I}}_{[t\!+\!1]} = \mathbb{H}_{GDL}(\mathbb{I}_{[t+1]}+f_{[t]},\Delta_0, \Phi_{[t]} )}
\end{equation}

%
% Interactive Hypergraph Model Exploration
%
\section{Interactive Hypergraph Model Exploration}
\label{ch:hypergraph_va}
In this section, we focus on the visualization and interaction with the temporal hypergraph model, providing a tight coupling between the data manipulation and display (see \autoref{fig:workflow}).
We begin by describing how the model state can be depicted using a matrix-based visualization that provides drill-down capabilities across multiple levels via semantic zoom. \changed{Drill-down is thereby defined as the seamless zooming through the different levels during exploratory analysis, starting from a general overview to increasingly more focused and detailed information, as highlighted in \autoref{fig:semantic_zoom_levels}.}
To facilitate the interactive exploration, we present user-steering based on classical filters for standard search tasks, a dynamically modifiable partition hierarchy to include user-based structuring, and various matrix reordering techniques for the focused analysis of connections and groups.
We then specify the interactions that allow domain knowledge to be incorporated into the machine learning model via relevance feedback and highlight how the updated predictions can be reflected in the existing visualization. This workflow facilitates the explainability of the underlying model, thus enabling the domain experts to provide more meaningful feedback. Finally, we describe how all interactions, domain knowledge input, and model output are stored in a provenance history, providing accountability and making the decision-making processes more transparent.

\subsection{Model Visualization \inlinesymbol{symbol_visualization}}
\label{sec:model_visualization}
\begin{figure*}
	\centering
	\includegraphics[width=\linewidth]{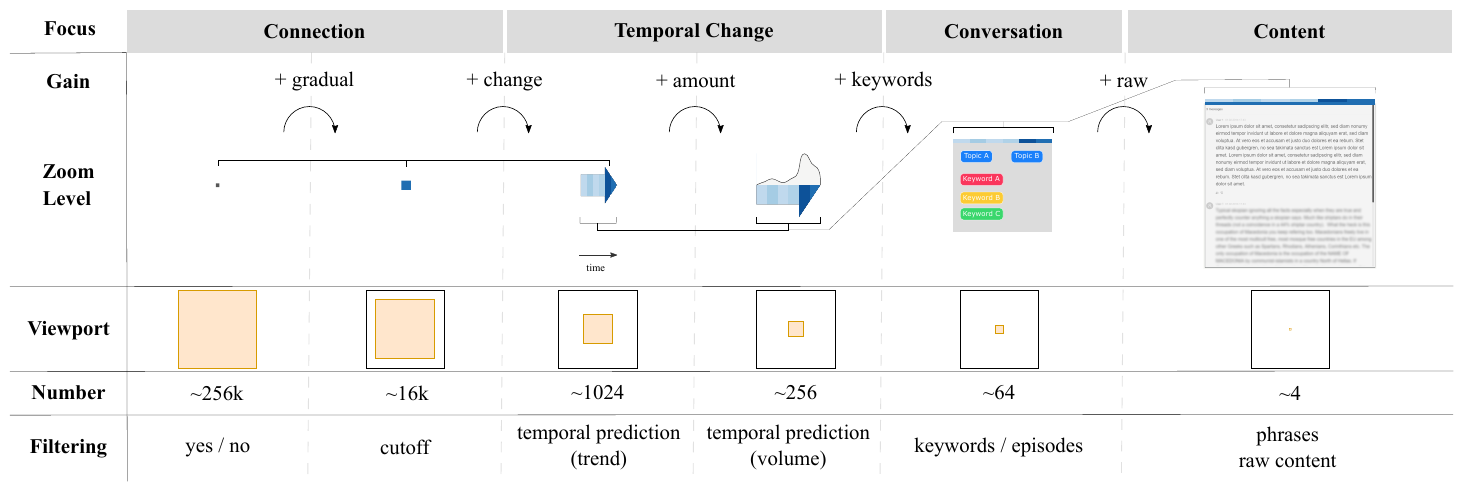}
	\caption{\updatenote{[Updated]} Semantic zoom levels and the different filtering levels\deleted{as described in detail in} (cf.~Section~\ref{sec:model_visualization}). 
		At each zoom step, the analyst gains another type of information about the model,\deleted{each step} filtering a different layer of complexity. 
		As the \changed{focus} becomes more detailed, the visualization takes up more space (zoom level and viewport as shown not to scale), while the number of visible entities decreases accordingly. \changed{The temporal predictions are shown in different forms throughout all levels (see fine grey line), with the detailed temporal evolution first shown in Level 3 and continuing down to Level 6.}}
	\label{fig:semantic_zoom_levels}
\end{figure*}
As discussed above, the complexity of temporal hypergraphs makes them difficult to visualize. 
Hence, we propose a multi-level matrix-based approach, specifically tailored to the hyper-dimensionality as well as the temporal component. 
The visualization (see \autoref{fig:teaser}) consists of a menu bar on top, controlling the interaction concepts discussed later, and, for the main part, a matrix-like viewport, showing nodes as rows and hyperedges as columns, with corresponding row and column headers. 
This viewport provides freely pan-able and zoom-able drill-down capabilities across six levels of semantic zoom, shown in \autoref{fig:semantic_zoom_levels}, increasing or decreasing the information detail: from an overview of model predictions down to contents.
\changed{For this purpose, we use three different level types: cells, arrows, and content boxes.
Colored cell visualizations are used in Levels~1 and 2.
An arrow-like representation reflecting a timeline is used in Levels~3 and 4.
The base of the arrow represents the past, while the head reflects predictions.
As the predictions become more uncertain with time, the arrowhead becomes smaller, reflecting the increased uncertainty and thus the decreased relevancy of the prediction.
Levels 5 and 6 add text-based elements like keywords or raw content.
Level 3 and beyond all contain the temporal aspect.}

The visualization depends on the zoom state of the viewport. 
During drill-down, the focus shifts from a general structure overview over the temporal evolution to the raw content, providing the expert with more and more detailed information.
\changed{Before we start with the description of this process, we define some necessary terms. As the feedback model outputs probabilities for the connections (see Section~\ref{sec:model_relevance_feedback}), gradual differences can be analyzed. When setting a minimum threshold for a connection to be meaningful, this allows for a \emph{binary choice}. Showing a color encoding of the connection strength allows for a more expressive representation of the \emph{gradual differences}. Setting a \emph{cutoff} threshold can still be used to avoid cluttering with low-probability entries.}
\changed{The drill-down shifts the focus of the analysis. It starts} at the \changed{(binary)} connectivity information, extends to gradual connection strength (Level~2), to the temporal change \changed{represented as an arrow} (Level~3), to the temporal change encoded using position instead of only color (Level~4), then to information summarizing the underlying content for the predictions, in this case, keywords (Level~5), and, at last, to the raw data (Level~6).\deleted{During this process the focus shifts. starting at (1) binary choices, to (2) cutoffs, to (3) temporal trends and (4) temporal trends expressed as volumes to (5) keywords, to the (6) raw content.}
\changed{The design choice for an arrow glyph representation in Levels~3 and~4 is based on five reasons: (1) The principal idea of an arrow glyph was previously published~\cite{Streeb.VAHypergraphPrediciton.2019} and found to be beneficial. Then, (2) given the target audience, a representation as an \emph{arrow of time} is closely related to everyday experience. Further, (3) the separation into arrow base and head allows a clear distinction between past data and model predictions, which is very important for the target audience. The arrowhead also allows to visually reflect the decreasing prediction accuracy by becoming smaller. In terms of (4) visual advantages, an arrow provides a distinct shape, while, e.g., a cell is easily perceived to merge with neighboring cells, which is undesired. The choice also comes with disadvantages, introducing white space and can sometimes lead to distracting patterns. Finally, (5) a design study on combining timeline and graph visualization by Saraiya et al.~\cite{Saraiya.VisGraphAssocTimeSeries.2005} shows that our approach---simultaneously overlaying the timeline---is best suited for detecting outliers. This is one of the main tasks for these levels, given the focus on change. The study also supports the design choice of showing only a single timestep in Levels 1 and 2, as the focus is on the topological structure. However, different visual representations like horizon graphs might be better suited when focusing on a continuous analysis.}
The \changed{seamless changes between levels} speed up navigating through large models while eliminating demanding context switches. 
Moreover, at each step, the information becomes more complex, requiring more screen space to visualize. 
For a regular HD screen, we give rough guidance ($\mathcal{O}(n)$) on the number of elements that can be usefully shown on-screen, \changed{amounting to around 256k grid cells of connectivity information and around four for the raw content.}
\deleted{Accounting for space taken up by the web client, as well as avoiding pixel-based representations due to burring issues in web environments, we estimate that around 256k grid cells of connectivity information can be shown, and around four for the raw content, allowing enough space for comparison.}

\subsection{Interactive Exploration \inlinesymbol{symbol_control} and Drill-Down}
To facilitate the interactive exploration, we contribute a user-steering based on classical concepts and filters for standard search tasks, a dynamically modifiable partition hierarchy to include user-controlled structuring and various matrix reordering techniques for the focused analysis of connections and groups.
All these interactions concepts are reactive, and the visualization can smoothly and instantly update ($<100$~ms), except for the domain knowledge integration in Section~\ref{sec:model_updates}.

\paragraph{Interaction and Filter Concepts \inlinesymbol{symbol_search}} Standard methods available in an interactive visualization are included, like (1) highlighting selected rows or columns, (2) highlighting hovered cells, (3) tooltip-based menus, (4) marking (i.e., starring) individual entries to highlight them for tracking and follow-up, (5) adding textual notes, and (6) showing additional meta-information. 
Modal views allow to (7) control the partition hierarchy (see details below), while setting an (8) overall cutoff threshold allows controlling the confidence threshold of the underlying model. 
A (9) global search function provides the ability to search for node- and edge information as well as content and highlights the matching components. 
At last, the menu bar allows (10) controlling the matrix reordering (see detail below).

\paragraph{Dynamically Modifiable Partition Hierarchy \inlinesymbol{symbol_hierarchy}}
\label{sec:partition_hierarchy}
To allow domain experts to articulate their mental categorization to the model, the experts can create (nested) groups of different nodes or hyperedges, creating hierarchies. \changed{The nodes or hyperedges hereby relate to the leaves of the dendrogram.}
\changed{The groups} can be expanded or contracted either directly from the \changed{node or hyperedge} headers, visually indicated by color, or by editing them inside the partition hierarchy viewer \changed{in a modal overlay}.
\changed{The viewer} shows a dendrogram-based representation with freely reorderable entries.
\changed{Each branch of this dendrogram can be independently collapsed or expanded, i.e., the abstraction level is local to each branch and not globally set. For example, it is possible to collapse a large, uninteresting sub-branch, including the nested nodes it contains, while simultaneously having one branch fully expanded and another only up to the penultimate level.} 
This is also independent of the overall visualization level, similar in concept to multiple fixed magic lenses, visually supporting different analysis paths.
The hier\-archy allows, for example, to group complementing entities together, to build meta-entities, and even hierarchies of entities.

\paragraph{Matrix Reordering and Sorting \inlinesymbol{symbol_gears}}
\label{sec:matrix_reordering}
To support the tasks relevant for our driving application (see Section~\ref{sec:tasks}), a matrix reordering is desirable such that related users and topics appear close to each other.
Due to the independent and often conflicting interpretations of both axes and the sparseness of the underlying matrix, the direct application of standard 2D numeric sorting algorithms (e.g., Multi-scale-, Chen-, or Travelling salesman problem ordering)~\cite{Behrisch.MatrixReordering.2016} often leads to unsatisfactory results, as they are mainly applicable to pairwise comparison matrices.
\begin{figure}
	\begin{subfigure}[t]{0.32\linewidth}
		\centering
		\includegraphics[width=\linewidth]{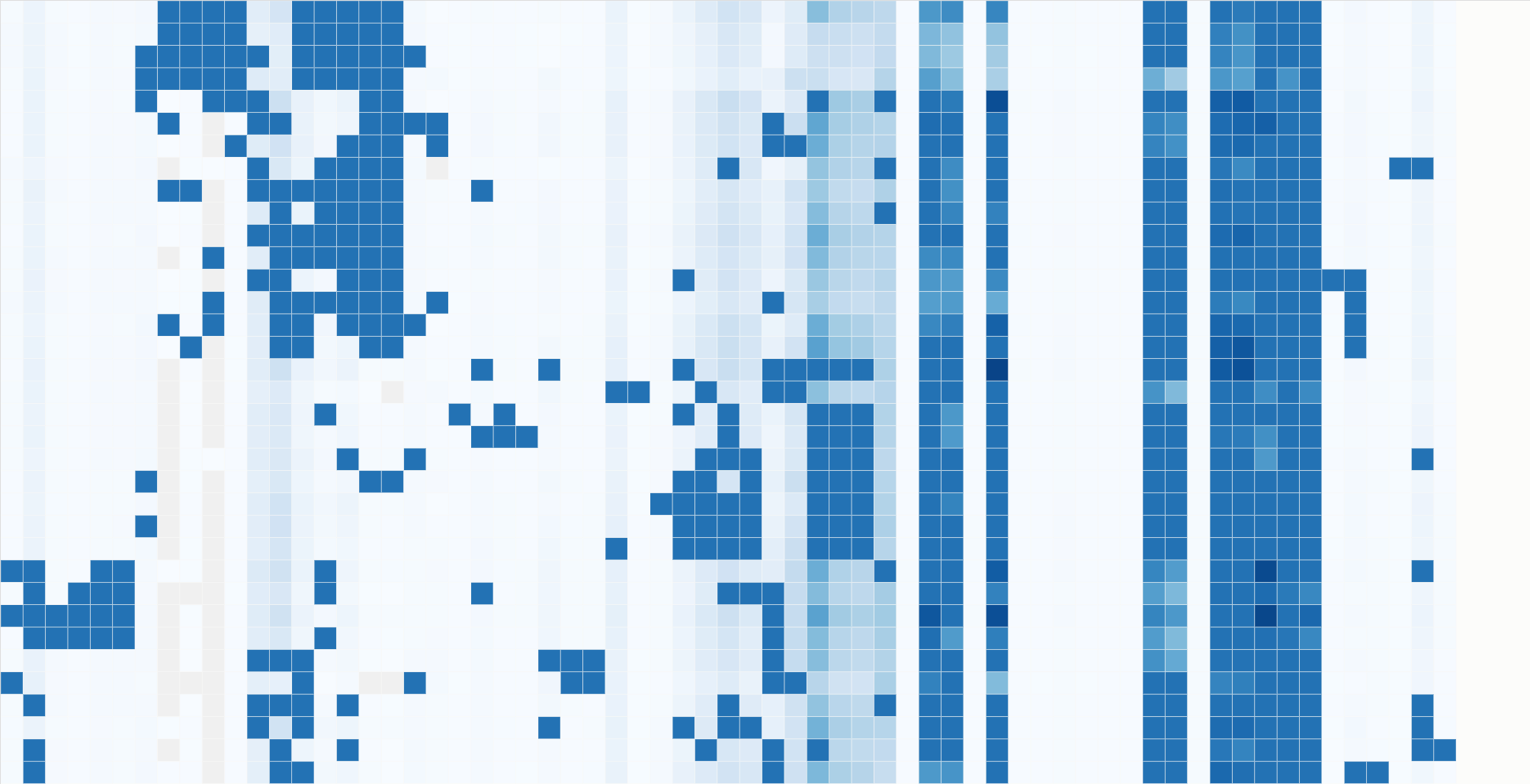}
		\caption{Multi-step hierarchy}
		\label{fig:matrix_reordering_behrisch}
	\end{subfigure}
	\hfill
	\begin{subfigure}[t]{0.32\linewidth}
		\centering
		\includegraphics[width=\linewidth]{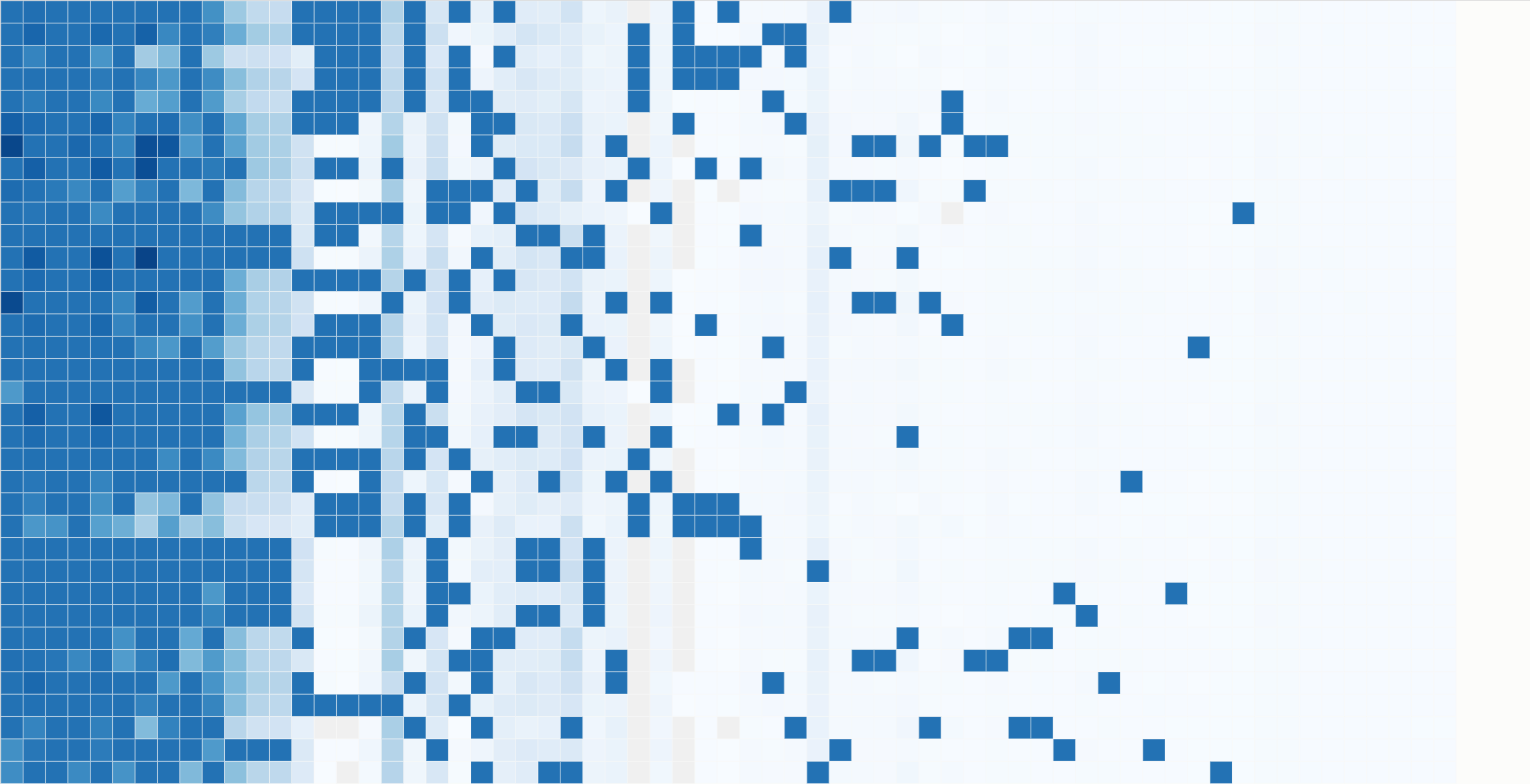}
		\caption{Size}
		\label{fig:matrix_reordering_size}
	\end{subfigure}
	\hfill
	\begin{subfigure}[t]{0.32\linewidth}
		\centering
		\includegraphics[width=\linewidth]{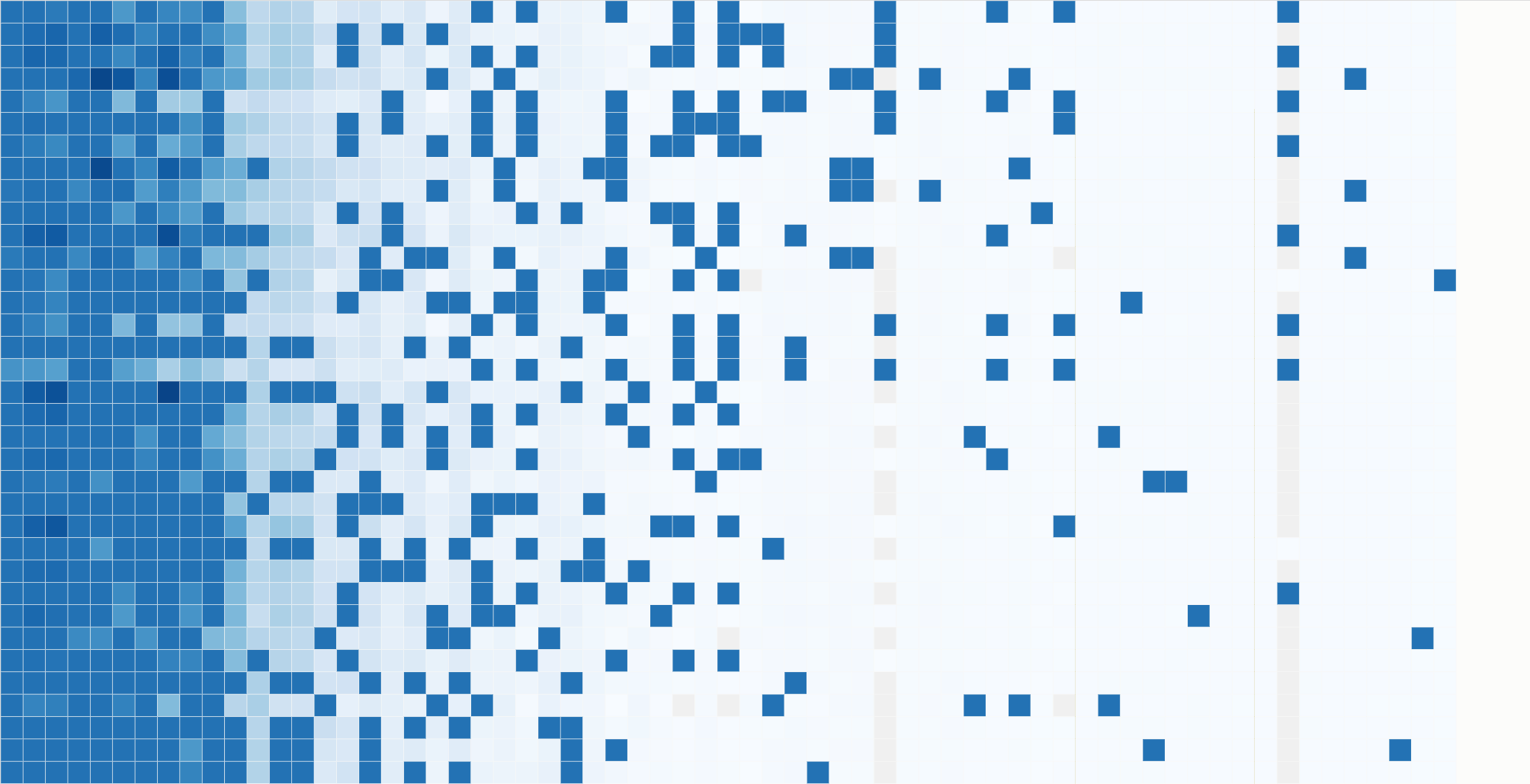}
		\caption{First occurrence}
		\label{fig:matrix_reordering_unsorted}
	\end{subfigure}
	\caption{\updatenote{[Updated]} Comparison of different matrix reordering techniques to facilitate the detection of similar groups and connections. Compared to the unordered state and the slightly improved ordering by size, the adoption of a \changed{default} multi-step, dendrogram-based reordering, modified and adapted from~\cite{Behrisch.MatrixReordering.2016}, enhances the clustering by similarity.}
	\label{fig:matrix_reordering}
\end{figure}

As part of the visualization, we offer three main different reordering strategies, as shown in \autoref{fig:matrix_reordering}: \changed{(a) matrix-reordering (default), (b) sorted by size (connectivity), (c) first occurrence (original).} 
\deleted{The latter is recommended for most use cases and comes in several different versions. It is used as a default.}
The reordering is applied individually for each axis, as the requirement may differentiate between search tasks, not always favoring a block-like clustering. % (see also Section~\ref{ch:discussion}). 
It also provides more flexibility for adopting other sorting methods in domain adaptions of our technique.
The underlying sorting principles build upon a dendrogram-based serial matrix reordering discussed by Behrisch et al.~\cite{Behrisch.MatrixReordering.2016}. 
It forms a multi-step process, combining the sorting of node and edge similarity vectors. 
Supported dendrogram methods are ward-, single-, average-, and complete linkage, combined with any pairwise distance function like Euclidean, cosine, or Jaccard. 
We refrain from discussing individual choices, which can vary strongly on domain adaption. 
For our case study, the Jaccard and cosine distance provide consistent results.

\subsection{Visual Analytics for Model Updates \inlinesymbol{symbol_provenance}}
\label{sec:model_updates}
\begin{figure}
	\begin{subfigure}[t]{\linewidth}
		\centering
		\includegraphics[width=\linewidth]{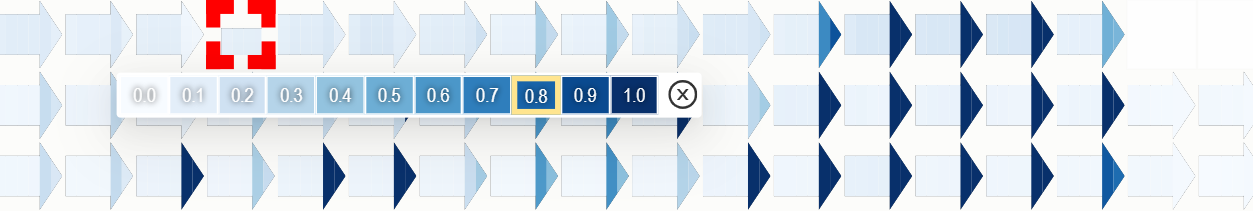}
		\caption{Manually input a connection strength.}
		\label{fig:change_feedback_visualization_input}
	\end{subfigure}
	\newline
	\begin{subfigure}[t]{\linewidth}
		\centering
		\includegraphics[width=\linewidth]{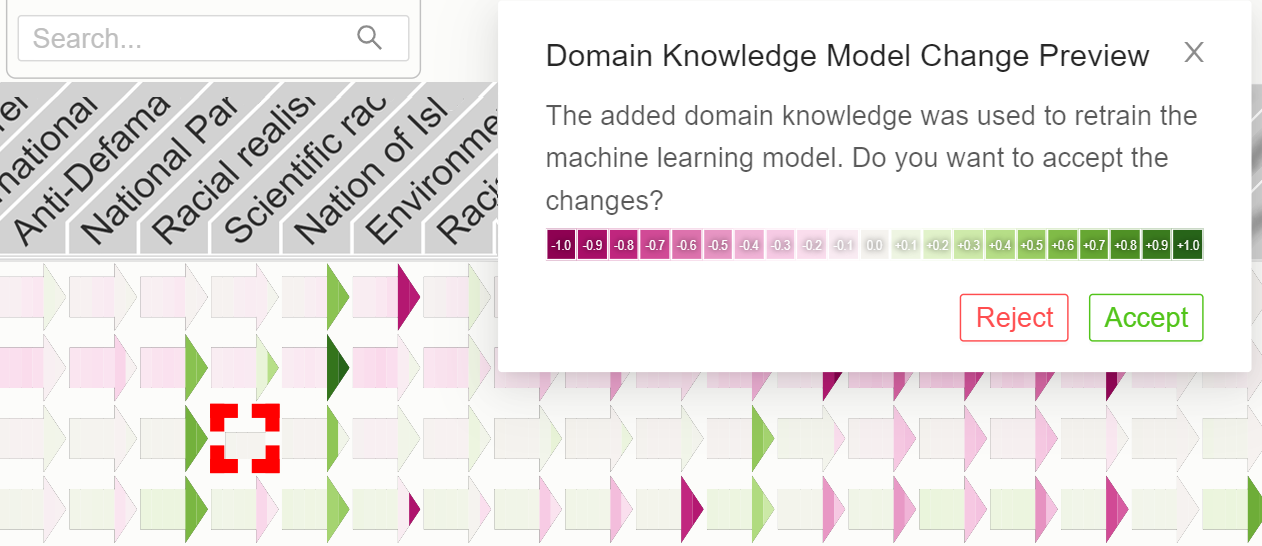}
		\caption{Resulting changes in the model predictions.}
		\label{fig:change_feedback_visualization_change}
	\end{subfigure}
	\caption{\updatenote{[Updated]} Resulting changes in the model prediction (ripple effect) from an \textcolor{ColorDKModelSelection}{input} are visualized by a diverging color scale (from \textcolor{ColorDKModelChangeNegative}{negative} to \textcolor{ColorDKModelChangeNeutral}{no} to \textcolor{ColorDKModelChangePositive}{positive} change). They can be explored and \textcolor{ColorDKModelChangeReject}{rejected} or \textcolor{ColorDKModelChangeAccept}{accepted}. This allows for model verification and multiple, different analysis paths.} 
	\label{fig:change_feedback_visualization}
\end{figure}

To increase the traceability of domain knowledge integration and explainability of the resulting model changes, we propose an interactive change feedback visualization, that seamlessly integrates with our vis\-u\-al\-ization. 
The two-step process is shown in \autoref{fig:change_feedback_visualization}.
An expert can integrate domain knowledge by selecting a cell and setting a new connection strength (\autoref{fig:change_feedback_visualization_input}), \changed{thereby complement missing or override model \emph{input} data.}
This input is used to partly retrain the model and refine its predictions as described in Section~\ref{sec:model_relevance_feedback}, leading to a ripple effect.
\changed{Thereby, the model has prediction authority, i.e., the user cannot manually fix the ultimate output to guarantee model authenticity.}
A spinner indicates the few seconds long operation. 
The resulting \emph{changes} are displayed inside the same view (\autoref{fig:change_feedback_visualization_change}). 
A diverging color scale is used, showing changes instead of predictions. 
Through two visually distinct scales, it is immediately apparent if predictions or changes in the predictions are shown. 
\changed{The view integration} allows for consistency, reducing the mental workload, and improving mental mapping.

Changes can be inspected on all levels of the visualization. The exploration is \emph{not} restricted to just the current viewport, finding even weak connections.
\changed{Change detection is facilitated}, allowing \changed{rejection} if deemed implausible or \changed{acceptance} if convincing, enabling \changed{the followup of} multiple analysis paths.
By\deleted{allowing to} iteratively and interactively queering the model and see how it responds to domain knowledge integration, experts can discern better how connections and processes in the model are related, improving understanding and increasing explainability.

%\subsection{Accountability by Provenance }
\changed{Experts in many applications} are interested in their analytical progress and \changed{must reproducibly document the steps.}
We address this by a re-loadable provenance, storing the interaction sequence, domain knowledge input, model output, and fixed RNG seeds.
This allows for inspection, verification, and traceability while providing accountability and making decision processes transparent.
The provenance history allows undoing analysis steps, \changed{preventing} dead-ends, revisiting and explaining past steps, but also bridging off\deleted{from a specific starting point for different,} to diverging analysis trails.

%
% Case Study
%
\section{Case Study: Internet Forum Communication Data}
\label{ch:case_study}
\changed{To demonstrate the visual exploration of temporal hypergraph models in \textsc{Hyper-Matrix}}, we conduct a case study, showing the applicability of our technique and improvements compared to existing approaches.\deleted{The first was implemented and assessed practically (cf.~Section~6), while the second describes a potential extension.}

\begin{figure*}[ht!]
	\begin{subfigure}[t]{.50\textwidth}
		\centering
		\includegraphics[width=.83\linewidth]{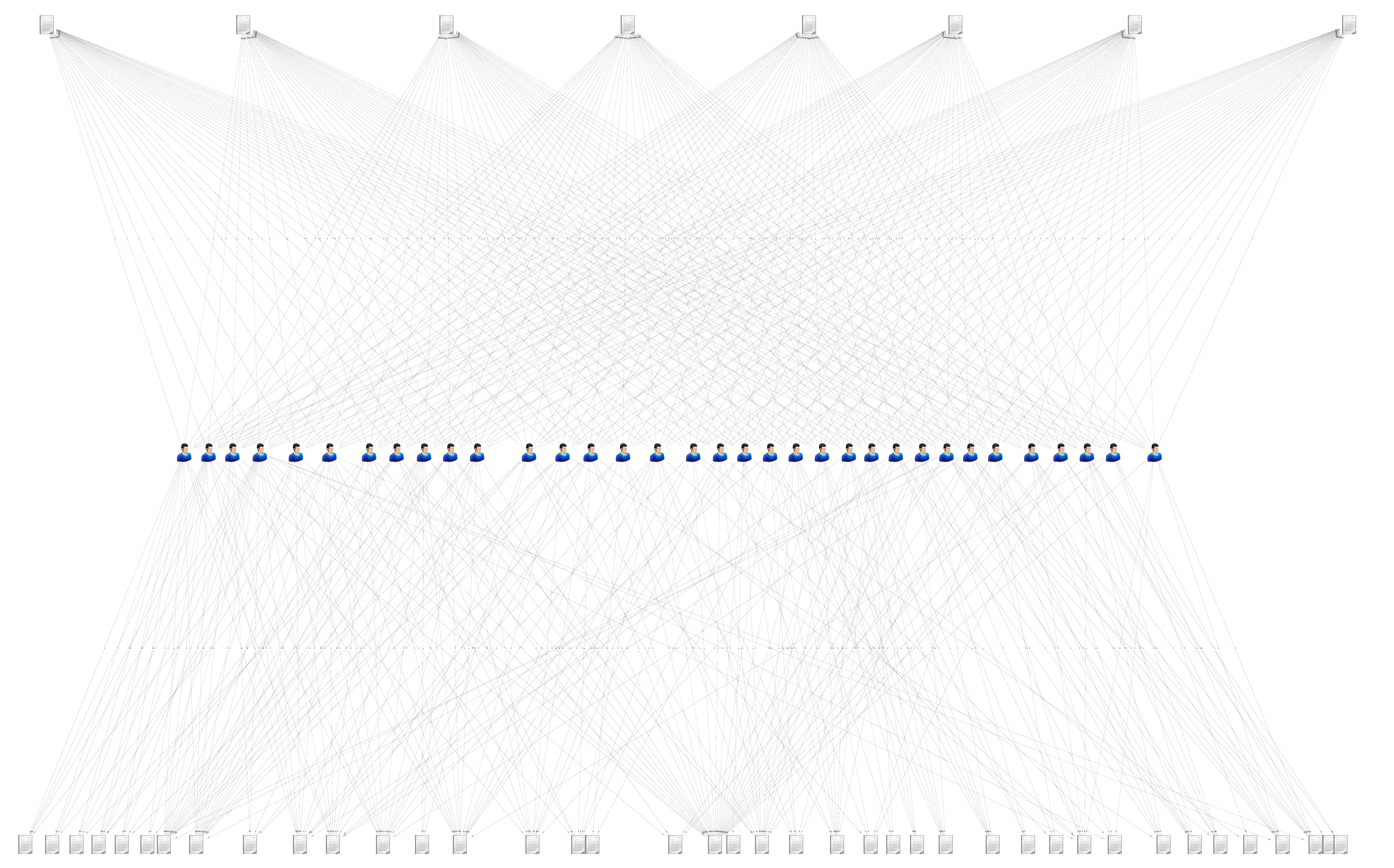}
		\caption{IBM i2 Analyst's Notebook. \changed{Automatically generated graph representation} from the hypergraph model displaying the connections \changed{(labels removed)} for the furthest predicted year using a modified bipartite representation. Data-wise, this can be compared to the connectivity information in our Levels 1 and 2. Clutter and occlusion prevent a meaningful global analysis, and while individual users and topics can be explored, this is only possible slowly, not without difficulty, and likely requires moving entities around to identify connections safely.}
		\label{fig:comparison_ibm}
	\end{subfigure}
	\hfill
	\begin{subfigure}[t]{.48\textwidth}
		\centering
		\includegraphics[width=0.7\linewidth]{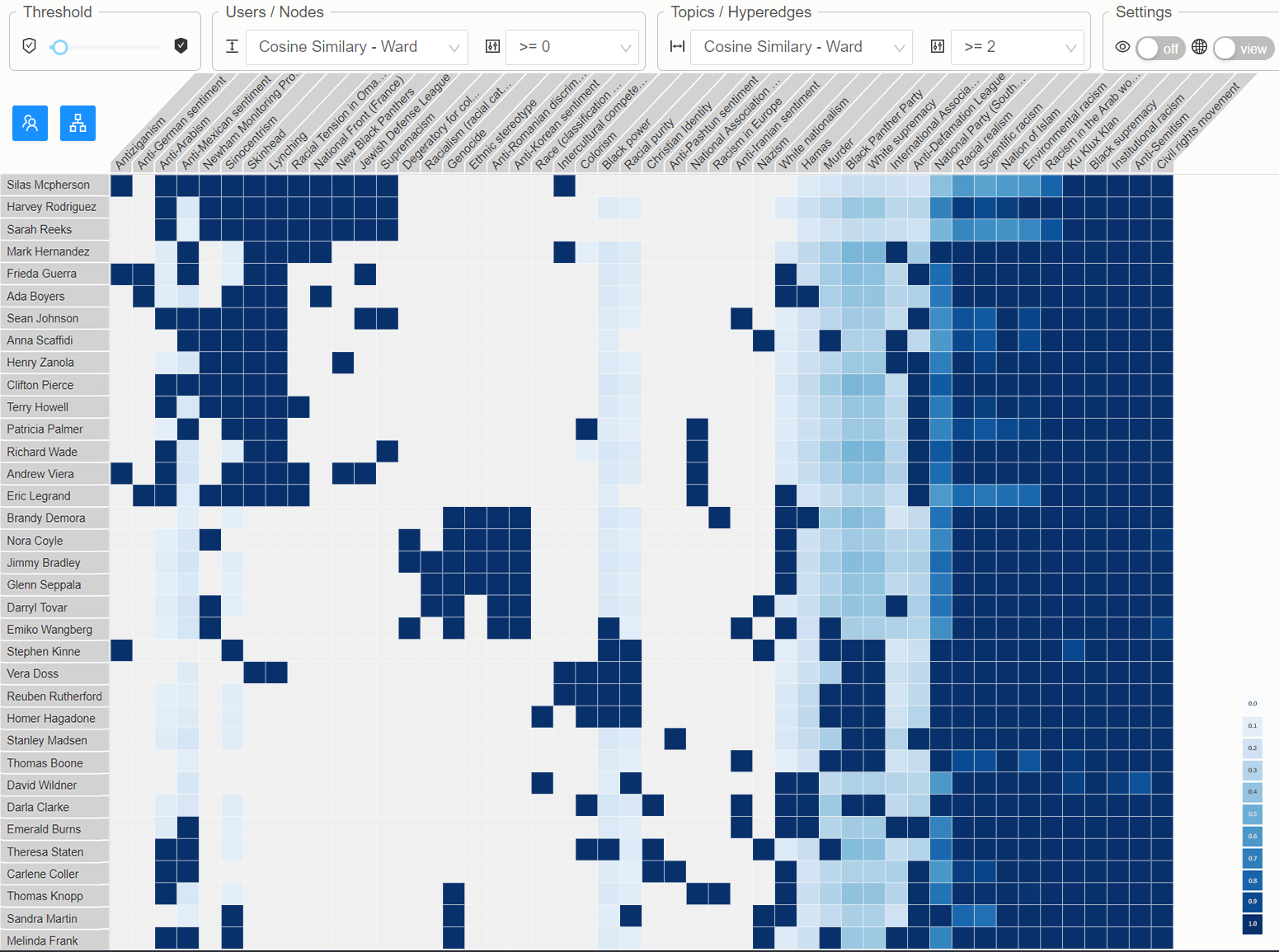}
		\caption{\updatenote{[Updated]} Our technique at Level 2, showing the same predicted \emph{connectivity} information as Analyst's Notebook in \autoref{fig:comparison_ibm}. Clusters and related users/topics can be pinpointed more easily. The color scheme and filtering settings in the top menu bar also facilitate to identify the prediction strength, \changed{which can be estimated by using the overlayed legend in the bottom right corner. The blue buttons allow to access the partition hierarchy modifier to view a dendrogram view of the grouped entities.}}
		\label{fig:comparison_our_level2}
	\end{subfigure}
	\newline
	\begin{subfigure}[t]{.50\textwidth}
		\centering
		\includegraphics[width=\linewidth]{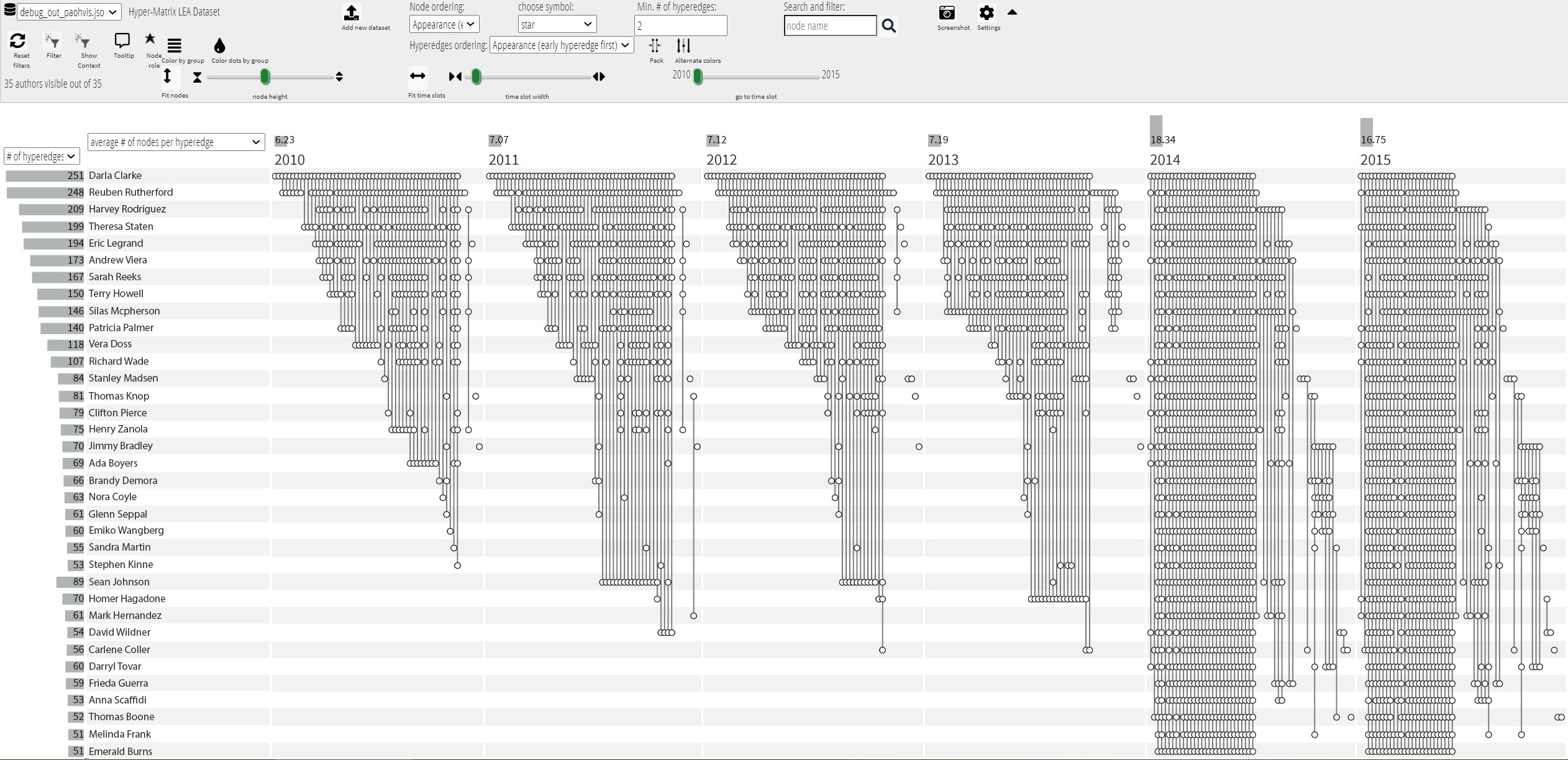}
		\caption{\updatenote{[Updated]}PAOHvis~\cite{Valdivia.DynamicHypergraphs.2018}. The temporal hypergraph evolution shows the individual hyperedges, allowing to find connected users and topics. However, the hypergraph size is at the upper limit for a feasible visualization, already leading to some cluttering. Also, due to the temporal splitting, the comparability between years is hindered for such complex, non-sparse hyperedges compared to our technique, \changed{but better suited for comparing topics in the same year}.}
		\label{fig:comparison_paohvis}
	\end{subfigure}
	\hfill
	\begin{subfigure}[t]{.48\textwidth}
		\centering
		\includegraphics[width=0.7\linewidth]{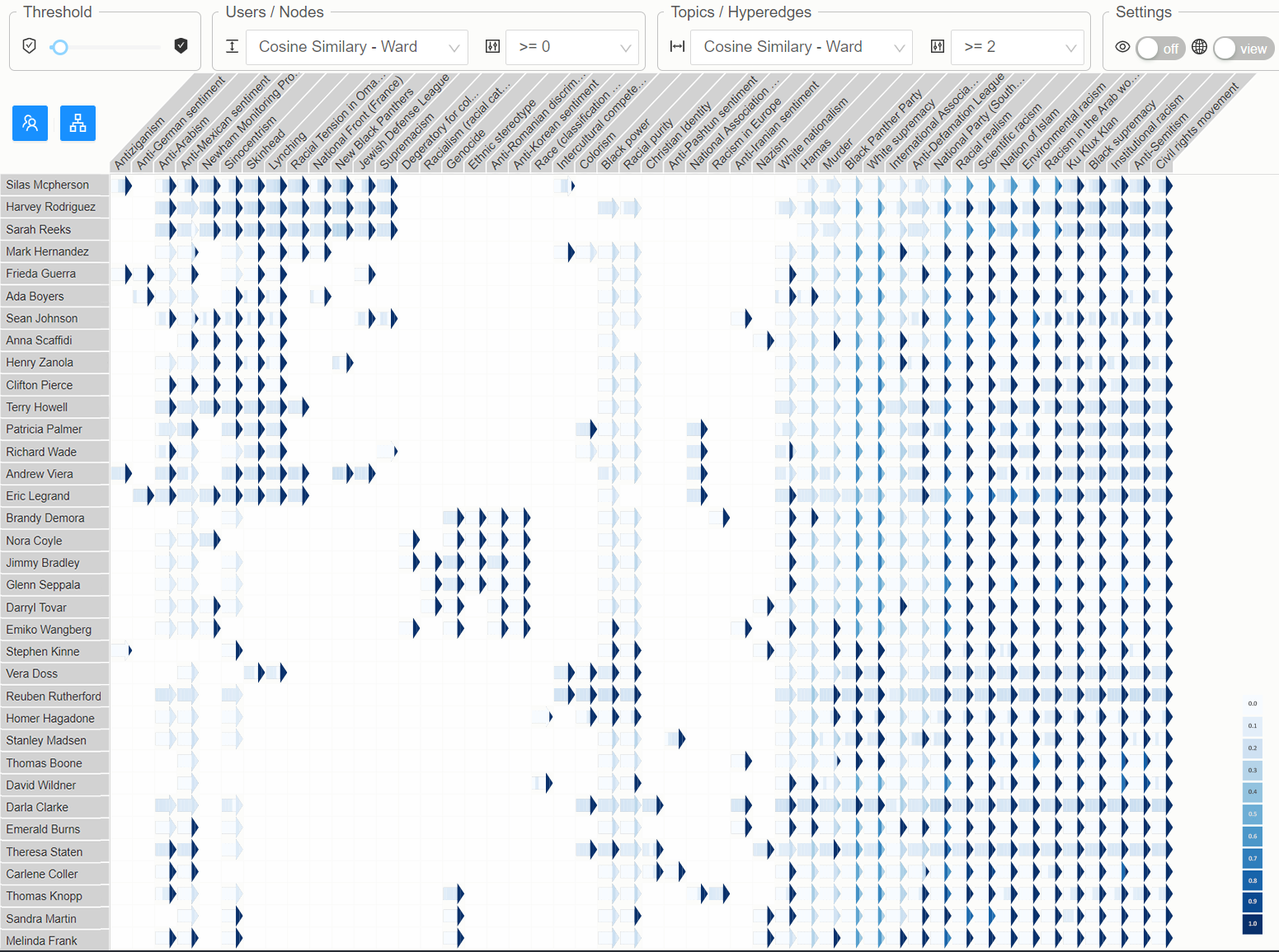}
		\caption{\updatenote{[Updated]} Our technique at Level 3, showing the same \emph{temporal evolution} information as \mbox{PAOHvis} in \autoref{fig:comparison_paohvis}. The scalability is increased, showing no occlusion and the comparability of trends \changed{(important for the case study)} is improved. \changed{This is due to retained cell ordering and short comparison distance. The downside is a reduced comparability between topics in the same year.} \changed{The nature of the predictions depend on the model.}}
		\label{fig:comparison_our_level3}
	\end{subfigure}
	\caption{Case study comparison of different approaches using the same internet forum hypergraph model dataset and exactly the same data view (connection strength $>0.1$, min. 2 hyperedges). Compared are the state-of-the-art industry solution IBM i2 Analyst's Notebook (\autoref{fig:comparison_ibm}), PAOHvis (\autoref{fig:comparison_paohvis}) against our technique, showing the information at two different levels of abstraction (Figures~\ref{fig:comparison_our_level2} and \ref{fig:comparison_our_level3}). Further, both external approaches only support a fixed network while our technique allows for an interactive refinement and domain knowledge integration.}
	\label{fig:comparison}
\end{figure*}

The communication data was collected from an internet forum well-known to law enforcement.
It contains $335\,188$ text posts from $4904$ users.
We pre-processed the data using standard NLP methods to extract $158$ topics, based on a domain-specific ontology. 
As described in Section~\ref{ch:methodology}, users are associated with nodes and topics\deleted{they talk about} become dynamic hyperedges.
To allow for a reasonable side-by-side comparison with the existing approaches, shown in \autoref{fig:comparison}, we had to restrict to a subset\deleted{within the above dataset}, consisting of 35 users, 65 topics, and six timesteps. This is around four times more than conventional approaches are designed for. We confirmed that our prototype works for significantly larger networks (cf.~Section~\ref{ch:discussion}).
Our prediction model is fed with \changed{four years (timesteps) of} historical data and then predicts the evolution of the \changed{next two years as two timesteps}.
Almost any real-world data is noisy and may miss some relationships.
Consequently, some of the conclusions drawn here may be inaccurate.
However, we focus on demonstrating the concepts and benefits of the visual analysis process \textsc{Hyper-Matrix} provides.

The task we want to focus on in this case study is the identification of related groups and missing links, common in criminal investigations.
\changed{To identify users discussing the same topics and topics discussed by the same group, the matrix reordering and connectivity information in Level 2 can be used to see structures, as shown in \autoref{fig:comparison_our_level2}. Their spatial closeness acts as primary identification criterion}, as similar row/column vectors are grouped closely. \changed{From this, their spatial closeness, describing the multi-step alignment, supports discovering related users or topics discussed simultaneously, but also latent connections.}\deleted{Compared to the industry standard Figure~6a, this view is faster and less error-prone to read.}\deleted{To find existing users by name or property, the search interface can be used.}\deleted{But also latent connections are easier found compared to a node-link diagram.}
Distinct orderings can be applied separately \changed{to nodes and hyperedges}, for example, to either favor overall similarity (cosine) or matching parts (Jaccard).
For other requirements, it is also possible to include different metrics.
To reduce noise and exclude weak connections, \changed{the top menu allows to set a threshold} for the connection strength for historical and predicted data.
\changed{A flag controls the ordering mode to either} respect the filtered or the full dataset (including filtered elements).
To further structure the view, the experts can manually \changed{click and select to} group users and topics to reflect their mental categorization of users and topics.
This allows to reflect domain-specific ontologies (e.g., similar concepts) or represent known formations of users.

\changed{Zooming to the lower visualization levels shows the} temporal development.
Compared to existing approaches (see \autoref{fig:comparison_paohvis}) our technique (\autoref{fig:comparison_our_level3}) increases the scalability and comparability for dense temporal evolution.
\changed{Compared to the industry standard }\autoref{fig:comparison_ibm}, presenting the \changed{temporal evolution}\deleted{intuitively} as a \changed{timeline-like} arrow within each cell reduces comparison distances.
\deleted{The two lowest levels}\changed{Levels 5 and 6} allow an expert to understand the actual data on which a predicted connection is based: The main keywords of the relevant text fragments\deleted{in Level 5} and\deleted{in Level 6}\changed{, respectively,} the actual raw text fragments (cf.~\autoref{fig:semantic_zoom_levels}). 
\deleted{This supports the expert in identifying issues; for example, irony or coded synonyms are still difficult to detect automatically.
	For this case study, we assume the expert has found an issue, but the same concept would apply if the expert has some additional information, either as external knowledge or as intuition about the analyzed group, and adds this knowledge.
	The model is immediately retrained, and a ripple effect can be observed, changing the overall prediction.}
\changed{This ability allows the expert to verify predictions and detect shortcomings as,} for example, irony and coded synonyms are still difficult to be detected automatically.
\deleted{For this case study, we assume the expert has found an issue, but the same concept would apply if the expert has some additional information, either as external knowledge or as intuition about the analyzed group, and adds this knowledge.
	The model is immediately retrained, and a ripple effect can be observed, changing the overall prediction.
	The feedback change preview allows the expert (cf.~Figure 5b), can correlate this with his intuition or facts, accept or reject the update, and continue with the exploration. }
\changed{If the expert has identified shortcomings on any level, e.g., missing connections or wrong attribution of an ambiguous term, the technique allows for the inclusion of this additional domain knowledge.
	To externalize knowledge, the expert selects the corresponding connection and specifies the proposed strength on a scale between 0 and 1.
	This translates to definite knowledge about no and guaranteed connection, respectively.
	More nuanced values like .7 allow the expert to reflect his own uncertainty.
	This allows them to try out hunches while simultaneously preserving some model flexibility.
	For this reason, the change preview} (cf.~\autoref{fig:change_feedback_visualization_change}) \changed{is extremely relevant for the domain experts, as it allows them} to see directly how their knowledge transforms \changed{the model prior to accepting the changes.
	They can explore the consequences by zooming and panning through all levels and} correlate their findings with their intuition or other facts.
\changed{If unsatisfied, they can go back.
	Otherwise, they can continue and repeat} this visual analytics loop\deleted{with rapid feedback} multiple times.
\changed{This rapid feedback supports} the expert in refining the model \changed{without being blind to the resulting consequences, but being able to control and explore the latest model state at all times}.
\changed{As the domain experts focus is on exploratory analysis the iterative refinement supports finding} connections and missing links\deleted{more clearly} \changed{faster.
With domain knowledge that is difficult to be integrated a priori, step-by-step changes are more understandable.}

%\subsection{Case Study B: Coronavirus Transmission Chains}
\deleted{Case Study B: Coronavirus Transmission Chains
	To describe the broader applicability of our technique, we present the sketch of a second possible use case in the medical and political domain.
	The idea of this scenario was brought up by one of the domain experts during the assessment described in Section~?%}\ref{ch:domain_expert_asessment}.
	It details the application of our temporal prediction model to explore and predict the spread of the coronavirus pandemic in a country, based on modeling of the transmission chain and the topology of movement patterns.
	The expert proposed to consider individual cases (or aggregates of cases in regions) as nodes and the connectivity and transmission chains as hyperedges.
	Transmission chains reflect the sequence a pathogen is spreading an infection trough a community, but despite their name, it has been known for some time that multiple, simultaneous pathogen introductions play a role~cite{Agoti.TransmissionPatterns.2017}, making non-linear modeling more precise.
	It has also been shown that hypergraph modeling better accounts for both the community structure and also the non-linear dependence of the infection pressure~cite{Bodo.EpidemicPropHypergraphs.2016}.
	As a priori, confirmed transmission chains are hard, and sometimes impossible, to trace on a large scale, geo-location, and information about whereabouts could be used as a more readily available alternative. 
	In the next step, the existing blueprint model would be replaced for one that predicts infections based on the connectivity and interactions in the past, reading machine-readable infection numbers. 
	The temporal nature of our technique would allow predictions about upcoming, accumulating transmission chains resulting in clusters and thereby identify regions most at risk.
	The matrix-reordering techniques would allow for lining up and concatenate infection chains; to preserve the order of time; it might be necessary to add a different metric; the thresholding would allow ignoring regions with $R_0 < 1.0$ (reinfection number), not contributing to exponential spreading. 
	Then the dynamically modifiable partition hierarchy can be used to cluster locations together, emulating their geo-locations or movement flows, better reflected by their topological relation than their geospatial position on a map.
	The different levels of visualization would allow keeping an overview, while simultaneously allowing to explore individual cases or regions, enabling an expert to understand how local transmissions transferred and where domain knowledge might be needed to improve understanding.
	This knowledge could, for example, be the occurrence of a large social event, to increase the connection strengths, or the quarantine measurements are going into effect, then reducing the connection strength.
	Using the feedback change preview, the expert can directly see how his choices influence the model predictions and correlate this with actual events or observations.
	Compared to geo-based or manually adjusted models, our approach could make model development more interactive and knowledge-based, allowing experts to trial different factors, approaches, and explore results in a single view.
	As a person infected by SARS-CoV-2 is not immediately contagious, with a mean incubation period of 6.4 days~cite{Backer.Covid19Incubation.2020}, an early warning system of future outbreaks leaves a window of opportunity of a few days to predict outbreaks and take mitigation measures for regions most at risk.}

\section{\changed{Formative Evaluation}}
\label{ch:domain_expert_asessment}
\changed{We performed formative evaluation sessions involving three domain experts (P1-P3).}
P1 is a criminal investigator working for a European law enforcement agency, having more than 30 years of experience, 20 years spent in digital and criminal investigations.
His expertise includes communication and network analysis, familiarity with commercial systems like IBM i2 Analyst’s Notebook\cite{IBM.AnalystsNotebook}, the graph visualization tool Gephi~\cite{Bastian.Gephi.2009}, as well as the large network analyzer Pajek~\cite{Batagelj.Pajek.1998, Batagelj.Pajek.2002}.
P2 works at the same agency in a different division, and has more than 20 years of experience in criminal investigations, specialized in group structure and content analysis.
P3 is a senior project lead at a governmental research institute, studying analytical raw data analysis for more than ten years.

\subsection{Study Procedure}
The \changed{formative evaluation} was conducted individually via remote screen sharing, taking about 60 minutes.\deleted{The assessment started with eliciting the domain expert's expertise, as described in the previous paragraph.}
\changed{For later review of these remote screen sharing sessions, they were recorded after receiving the formal consent of the experts.}
\changed{In the first 10 minutes a demo presented how to perform the visual analysis, explore and refine data and processes, and integrate domain knowledge in the search process and in the machine learning model.}
\deleted{The tasks that were demonstrated to and then tried out by the experts while commenting on it}\changed{The next 30 minutes were spent between the experts using the system and providing feedback, as well as additional on-demand demonstrations. The tasks the experts performed include overview, the identification of the most promising leads, and the drill-down through the different zoom-levels down to the actual raw content, in this case, communication data.}
Further, we demonstrated and debated the different interaction techniques, like cutoff values and thresholds, matrix sorting and reordering strategies, and the dynamically modifiable partition hierarchy, as well as the machine learning feedback process.

\changed{In the last 20 minutes, the authors interviewed the experts asking 32 prepared questions (see annexes).}
\deleted{During the whole first part of the evaluation, taking around 40 minutes}\changed{During each of the formative evaluation sessions, the experts engaged actively, trying out concepts, asking questions, commenting on the features, and pointing out issues. If an expert already partially gave comments during the 30 minutes session, they were offered to extend their answer.}
For example, when an aversion or surprising idea was mentioned, we additionally focused on these aspects.
The interview was designed to elicit aspects of our technique that the experts \changed{find relevant for their work} or confusing or misinterpretable, as well as opinions on the individual approaches.
\deleted{We describe the findings of our formative evaluation below.}

\subsection{Findings and Lessons Learned}
The \textbf{main observations} during the study are that our approach can effectively support most analytical requirements of the experts and that the experts favor both the  \changed{\textbf{rapid exploration}} of large datasets at different levels as well as the ability to integrate and contribute with their  \changed{\textbf{domain knowledge}}. \changed{This matches with their need to identify general trends in single combinations of users and topics and simultaneously identify co-occurrences. For this, the general prediction is more important than being able to identify differences between entities in the same year (cf.~\autoref{fig:comparison}). The underlying model we built upon~\cite{arya2019predicting} has proven to perform sufficiently well in this prediction task with an AUC (area under curve) of the ROC (receiver operating characteristic) of $.88$ and a recall value of $.81$.}
Excluded from the requirements are concepts outside the design scope, like purely mathematical capabilities as, for example, general centrality calculations, for which algorithms exist and could be included.
In the following, we structure and summarize the main findings based on the expert's interactions and comments.

The domain experts agree that our approach of structuring information in  \changed{\textbf{multiple levels}} of details, using a  \changed{\textbf{matrix-based approach}}, is novel and therefore is not used in practice in their domain.
For example, so far P1 has worked with either text-based or graph-based tools, and thinks our approach can \enquote{perfectly complement} existing workflows. 
The experts highlight the ability to \deleted{so much information in such an easy way (P3)}\changed{effortlessly explore so much information (cf.~P3)}, thereby \enquote{saving time} (P1), enabling a \enquote{quick analysis} (P3), while providing a \enquote{great overview \dots\ with much details, \dots\ but without overloading} (P2) the analyst, with an ease that is \changed{unexpected, given previous experience with this amount of data (cf.~P2)}.
\changed{We observed, that the experts often switch between the levels for targeting (upper levels) and then exploration and confirmation (lower levels).}
As P2 notes, this increases the size limit of the visually analyzable graph models, enhancing upon existing systems. \enquote{Together with the search capability} (P1), this allows for a very flexible workflow, enabling a good overview even for larger datasets.

The initial overview visualizations (Levels 1 and 2) are welcomed for providing a fast overview (cf.~P1).
The \changed{\textbf{color scheme}} in Level~2 is regarded as \changed{comprehensible without explanation and aligning with expectations (cf.~P1)}.
It helps to provide guidance \enquote{where to start}~(P1), and \deleted{useful for their work (cf.~P3)}\changed{supports analysts in \enquote{planing their actions} (P3)}.
To make the color scheme absolutely comparable, P3 requested the addition of a color legend.
The  \changed{\textbf{glyphs}} are appreciated for providing details on the temporal distribution and future predictions (P2, P3).
The glyph-based arrow representation in Levels 3 and 4 is appreciated for providing details on the temporal distribution (cf.~P2, P3) interesting to the experts, and, most importantly, \enquote{the future predictions}~(P2) in context of the historical data. 
Depicting future predictions in the arrowhead and the past data in the shaft, and seeing both together was described as \enquote{helpful} (P3).
The alignment by fixed timesteps, like years, is regarded as precise and practicable (cf.~P1) by the experts.
In comparison, the distribution as line chart in Level 4 received mixed responses, with P1 and P3 finding it \changed{beneficial for their understanding} to get a better, absolute reading, while P2 feels \enquote{it does not add much}.
The  \changed{\textbf{keyword visualization}} (Level 5) is regarded as fine for an abstract summary of the content visualization but could be extended (cf.~P3).
This layer, representing the \enquote{main connection} (P1) to the actual raw data, is important (cf.~P1), and only shown when relevant in high zoom levels, \enquote{where the text content is relevant} (P1).

The ability to  \changed{\textbf{search}} through all underlying textual data and highlight matches in the views was received enthusiastically by all experts, as they can also transfer and fulfill some of their existing workflow, e.g., content- and text-based workflows, with our technique.
It allows to explore global tendencies while enabling to query locally (cf.~P2), not being distracted by other matches \enquote{not relevant at the moment} (P2).

While the visualization alone helps them already some ways, providing them \enquote{with improved degree of detail \dots\ unknown so far} (P1), all the experts also agree that the  \changed{\textbf{interaction concepts}} constitute an essential and relevant part of the approach, \enquote{helping them with strategical and operational decision} (P1).
The  \changed{\textbf{matrix reordering strategies}} significantly improving the visual clarity of the overview, are regarded as \enquote{very interesting} (P2), and enable the experts to detect \enquote{groups} (P1) as well as connections easily, allowing them to \enquote{quickly identify hotspots} (P2), while putting less emphasis on weak connections.
This is regarded as \changed{very supportive, being rarely supported in analysis systems (cf.~P1),} \enquote{saving costs and time} (P1).
\changed{We observed that the experts use this as system guidance.}
The  \changed{\textbf{partition hierarchy}} is regarded by all experts as \enquote{essential} (P2), with P3 describing it as a \enquote{core functionality}. 
It allows grouping different model parts into physical concepts\deleted{groups}, applying structure comparable to existing mental models (cf.~P2), improving the mental mapping.
It \enquote{makes decision easier} (P3) and allows to \enquote{connect things} (P3).

The experts further describe that with existing tools, one major problem is that their  \changed{\textbf{mental concepts}} and models can \enquote{not [be integrated] enough} (P2) in the exploration, making it harder and less \changed{comprehensible}.
They notice that our approach supports them in three ways not present in existing tools: (1) the  \changed{\textbf{interactive exploration}} allowing to follow their instinct, (2) the modifiable  \changed{\textbf{partition hierarchy}} to express and capture their mental concepts, and, \enquote{most importantly} (P1), \changed{(3)} the ability to integrate their  \changed{\textbf{domain and external knowledge}} directly in the model.
While the experts wished that they could already \enquote{generate a report [$\ldots$ and] export single entries} (P2) as commercial systems do, they note the enormous conceptional benefits of our technique.
They regard them as \enquote{optimal} (P1), as there \enquote{are concepts and knowledge that cannot be modeled with machine learning [alone]} (P1) and are not \enquote{available} (P1) in the data.
This knowledge then \enquote{cannot be integrated so far} (P1), is often documented in the head of the domain expert or \enquote{on a post-it note on the desk} (P1), leading to a high risk of the knowledge being \enquote{lost} (P1) or not leveraged.
\changed{According to P1, the \textbf{knowledge integration} is performed iteratively during exploration, \changed{which we also observed as the experts adding knowledge intermittently,} beginning with their main suspects and then expanding, adding knowledge when necessary either from post-its or when reading a name triggers a memory.}
The experts think that our \changed{\textbf{feedback loop}} \deleted{highly useful (P1)}\changed{contributes to their analysis (cf.~P1)}, replacing and \enquote{perfectly complementing} (P1) existing workflows.
\changed{They regard the ability to \emph{interactively} insert their knowledge as versatile. P1 noted that inserting all knowledge beforehand would be error-prone and \enquote{practically impossible} for larger datasets.}
\changed{To see \enquote{\textbf{validation} [possibilities] on changes} (P3) is especially important for vetting, and the change view is regarded as \enquote{very clear} (P1), allowing them a first glance, beneficial for prefiltering, steering and follow-up search guidance (cf.~P1) to better divide their time for exploration. For improved usability P1 suggested to enable clicking to jump directly to the raw data in the change preview mode for validation.}
\changed{P1 regards the ability for a \textbf{global accept/reject} as sufficient for now, conceding that a partial accept could be explored in the future, although he does not see an immediate benefit.}
\changed{They state that the \textbf{0--1 scale} is \enquote{understandable and usable} (P1), but note that using the \enquote{5x5x5 system} (P1)---a commonly used police system based on letters A--E and 1--5 for source and intelligence evaluation~\cite{HMRCManual.555Model}---would be immediately understood and universally accepted in the target domain.}
The approach allows them to integrate their domain knowledge on multiple levels, together with the ability to perform a \enquote{quick analysis} (P3) of \enquote{large amounts of information} (P2) \enquote{in a targeted} (P1), non-overloading manner.
\changed{From the observations of the experts, we derived a set of \textbf{tentative tasks}, relevant in law enforcement: (1) finding linked users/topics, (2) connecting users which share related topics to identify co-conspirators, (3) using classical text-based search in the raw data to identify users, (4) finding and judging an in/decrease of user activity for a topic, (5) finding a temporal co-occurrence between topics and users, (6) adding domain knowledge to a specific user and specific topic and judging the implications, (7) transfer raw data patterns and identify related users, and (8) confirming the model predictions by cross-validation plausibility with the raw data texts.}

\section{Discussion and Future Work}
\label{ch:discussion}
\deleted{As demonstrated above, there is considerable need for an interactive, intuitive visual analytics system for the previously described tasks.}
\deleted{Our approach provides the foundations for the exploration and refinement of temporal hypergraph models while including a feedback loop to integrate domain knowledge not only during the exploration but also in the model itself, thereby supporting the domain experts in their work more efficiently than previous approaches.}
During the evaluation, we received multiple proposals on how our approach could be extended further\changed{, including by mathematical analysis methods and industry-grade interfaces.}\deleted{and applied to different domains, posing interesting questions regarding extensibility and transferability.
	Some are due to the prototype nature of our approach, like too scientific terms in the interface (P2), requesting additional exporting capabilities (P1), or mathematical analysis methods (P3).
	Our technique is not a finished or commercial system, with the prototype implementation aligned along with the needs of the case study, the exploration of communication pattern evolution in the law enforcement field.}
In the following, we discuss the limitations and broader applicability of our approach, also in the context of future work.
For our prototype approach, we adapted the generic blueprint of a machine learning model to the case study. 
This use case has its own limitation, requiring structured data with time and author information, and dependent on advanced topic extraction models.
We tested our prototype successfully with $1\,000$ users, $800$ topics, and 15 timesteps on an HD screen, typically the upper size for large investigations.\deleted{Thereafter our current SVG-based rendering architecture becomes inefficient. The usage of WebGL would improve the performance and overcome this engineering limitation.}\deleted{At an even larger scale our rendering architecture would benefit from more direct GPU support.}
\changed{In terms of data type, the technique can cope both with sparse and none-sparse matrix structures. For the former, the matrix reordering allows to prioritize more relevant connections and order them further on the top left, reducing the required screen usage for the main parts. Of course, a \emph{homogeneous} sparse matrix does not benefit from that. In this case, and for none-sparse matrices, the different zoom levels shift the size limitations. Nevertheless, they do not scale infinitely.}
\deleted{Regarding to the general technique,}Scrolling would be needed when scaling further, even for the overview level.
\changed{According to domain expert P1, there the primary concern would be the number of users (y-axis), but using the partition hierarchy and matrix reordering could partially mitigate the issue. When increasing the number of time steps, the arrow becomes more detailed, shifting from blocks to a more continuous stream, becoming less distinguishable. For our use case, this fine-grained time is not primarily relevant because the experts aim at seeing who has recently been interested in a topic. However, it might become an issue when the task requires to extract detailed timestamps. Therefore one could use hovering, magnification on demand, or a more specialized visualization.}
\changed{Also, the visualization presented is better at analyzing trends and connectivity tasks on an overview level. Comparing the same time step in Levels 3 and beyond between two non-aligned nodes, however, becomes harder. For further work, we envision an adaptable overview layer showing a specific time point, allowing cross-cell comparability.}
When adapting to different use cases, some of the filtering methodology likely has to be changed.
For example, when supporting biochemical process analysis, the raw attributes are not texts anymore, which (1) would need a different visualization for the content in the two lowest display level, but would also impact (2) the search functionality, which would need to be adapted to search and filter for biological and chemical properties instead of text.\deleted{Further, currently, there is no standardized import and export data format, requiring the creation of custom reader and writer methods for different datasets.}
The discussed visualization components serve only as examples for the visual analytics workflow presented. 
When adapting to a different field, there exist manifold possibilities for extensions, by integrating domain-specific visualization components. 
We provision this by a modular view architecture, supporting independent layer modules.
\changed{Further enhancements are multiple magic lenses to allow for simultaneous drill-down to different levels.}\deleted{The visualization could be further enhanced by supporting zoom via multiple magic lenses for the matrix-view, allowing simultaneous drill-down to different levels, but complicating rendering.}

In the future, we envision improvements to the feedback system, for example, showing how domain knowledge propagates not only between two model states, i.e., before and after adding knowledge but also\deleted{being able to } explaining the effects of previously introduced knowledge, for example, by interactively highlighting the individual influences on hover.
\deleted{Due to the architecture of our approach, we can support this use case computationally and visually without significant problems.
	However, the computation time to consider each domain knowledge input separately for every state individually leads to a linear increase in model change computation, dependent on the number of domain inputs. This}
\changed{This is supported by our architecture, but the computation time scales linearly with the number of domain knowledge inputs, which} leads to computation times of several minutes and more, making it infeasible in an interactive environment for fast iterations.
We hope to improve this by enhanced engineering, reducing the model setup and reloading times by advanced ways of updating the hypergraph model.

\deleted{
The expert interviews provided very positive feedback for the workflow and the prototype. 
Yet, we are well aware that the study group was relatively small, given the problematic situation of performing user studies at the beginning of 2020, especially so in the target domain.
To broaden the perspective, as future work, we plan to evaluate with an open dataset. Further, the study could be extended with the tentative set of comparable tasks and undertaken with more participants. }

\section{Conclusion}
\label{ch:conclusion}
% The issue
Many processes are difficult to describe using traditional graph-based concepts and benefit from more precise yet more complex modeling as temporal hypergraphs.
We address this challenge by using a geometric deep learning approach and extend it to hypergraphs. 
However, such deep learning models typically do not incorporate domain knowledge, usually unavailable in the data. 
This is not least because domain experts struggle to articulate their knowledge without rapid, iterative feedback and intuitive representations matching their mental models, alternatively requiring a detailed a priori understanding of the problem. 
Hence, domain expertise is often not leveraged to its full potential.

% the solution
We contribute a technique, named \textsc{Hyper-Matrix}, to make temporal hypergraph model exploration more accessible for domain experts by enabling the integration of domain knowledge into the process and support their mental models through a multi-level matrix-based visualization architecture.
The technique enables the interactive evaluation and seamless refinement of such models while providing a tight coupling and rapid, iterative feedback cycles to the underlying machine learning model. 
Model changes in response to the integration of domain knowledge are visualized transparently by a change preview, allowing experts to foster a more detailed understanding of how the underlying model works while externalizing their knowledge to teach the machine.

The approach allows to swiftly explore vast search spaces while maintaining focus and eliminating demanding context switches. 
Drill-down capabilities across multiple levels allow studying details and model contents on demand while retaining the overview. 
This architecture facilitates a focused analysis of relevant model aspects, allowing experts to detect patterns more rapidly and accurately. 
It is complemented by interactive filtering and search, various matrix reordering techniques, and a dynamically modifiable partition hierarchy, allowing the integration of domain knowledge in the visualization layers.

% evaluation
We evaluate our approach in one case study and through \changed{formative evaluation} with law enforcement experts using real-world communication data.
The results show that our approach surpasses existing solutions in terms of scalability and applicability, enabling the incorporation of domain knowledge and allowing fast and targeted search-space traversal.
% model agnostic
While we focused on topic prediction for law enforcement as driving application, the interactions and concepts work with any temporal hypergraph, being model agnostic and applicable more generically to a wider variety of domains.
% final takeaway message
With our technique, we hope to pave the way for domain experts to a more interactive exploration and refinement of temporal hypergraph models, enabling them to use their knowledge not only for steering but also to articulate it into the machine learning model.

\acknowledgments{
	This project has received funding from the European Union's Horizon 2020 research and innovation programme under grant agreement No.~700381. This material reflects only the authors' views, and the Commission is not liable for any use that may be made of the information contained therein.
}

\bibliographystyle{abbrv-doi}

\bibliography{references}
\end{document}